\documentclass[12pt,letter]{article}
\pdfoutput=1
\usepackage{graphicx, epsfig, color,cite}
\usepackage{amsmath}
\usepackage{amssymb}
\usepackage{float}
\usepackage{caption,subcaption,graphicx}
\usepackage{hyperref}

\def\to{\rightarrow}

\def\bi{\begin{itemize}}
\def\ei{\end{itemize}}

\def\tchi{\tilde\chi}

\def\tb{\tilde b}

\def\tst{\tilde t}
\def\ttau{\tilde \tau}

\def\tg{\tilde g}

\def\alt{\lesssim}
\def\agt{\gtrsim}
\def\be{\begin{equation}}  
\def\ee{\end{equation}}  
\def\bea{\begin{eqnarray}}  
\def\eea{\end{eqnarray}}

\begin{document}
\begin{titlepage}
\begin{flushright}
OU-HEP-220731
\end{flushright}

\vspace{0.5cm}
\begin{center}
  {\Large \bf Prospects for heavy neutral SUSY Higgs scalars\\
    in the hMSSM and natural SUSY at LHC upgrades
%    and FCChh
    }\\
\vspace{1.2cm} \renewcommand{\thefootnote}{\fnsymbol{footnote}}
{\large Howard Baer$^{1,2}$\footnote[1]{Email: baer@ou.edu },
Vernon Barger$^2$\footnote[2]{Email: barger@pheno.wisc.edu},
Xerxes Tata$^3$\footnote[3]{Email: tata@phys.hawaii.edu} and
Kairui Zhang$^2$\footnote[3]{Email: kzhang89@wisc.edu}
}\\ 
%and Dibyashree Sengupta$^1$\footnote[4]{Email: Dibyashree.Sengupta-1@ou.edu}
\vspace{1.2cm} \renewcommand{\thefootnote}{\arabic{footnote}}
{\it 
$^1$Homer L. Dodge Department of Physics and Astronomy,
University of Oklahoma, Norman, OK 73019, USA \\[3pt]
}
{\it 
$^2$Department of Physics,
University of Wisconsin, Madison, WI 53706 USA \\[3pt]
}
{\it 
$^3$Department of Physics and Astronomy,
University of Hawaii, Honolulu, HI 53706 USA \\[3pt]
}

\end{center}

\vspace{0.5cm}
\begin{abstract}
\noindent
We examine production and decay of heavy neutral SUSY Higgs bosons
$pp\to H,\ A\to \tau\bar{\tau}$ within the
hMSSM and compare against a perhaps more
plausible natural supersymmetry scenario dubbed $m_h^{125}({\rm nat})$
which allows for a natural explanation for $m_{weak}\simeq m_{W,Z,h}\sim 100$ GeV
while maintaining $m_h\simeq 125$ GeV.
We evaluate signal against various
Standard Model backgrounds from $\gamma ,Z\to\tau\bar{\tau}$, $t\bar{t}$
and vector boson pair production $VV$.
We combine the transverse mass method for back-to-back (BtB) taus along
with the ditau mass peak $m_{\tau\tau}$ method for acollinear taus
as our signal channels.
This technique ultimately gives a boost to the signal significance over
the standard technique of using just the BtB signal channel.
We evaluate both the 95\% CL exclusion and $5\sigma$ discovery reach in the $m_A$ vs. $\tan\beta$
plane for present LHC with 139 fb$^{-1}$, Run 3 with 300 fb$^{-1}$ and
high luminosity LHC (HL-LHC) with 3000 fb$^{-1}$ of integrated luminosity.
For $\tan\beta =10$, the exclusion limits range up to
$m_A\sim 1$, 1.1 and 1.4 TeV, respectively.
These may be compared to the range of $m_A$ values gleaned from a statistical
analysis of the string landscape wherein $m_A$ can range up to $\sim 8$ TeV.
\end{abstract}
\end{titlepage}
%\pacs{12.60.-i, 95.35.+d, 14.80.Ly, 11.30.Pb}
%12.60.-i   Models beyond the standard model
%95.35.+d   Dark matter

\section{Introduction}
\label{sec:intro}

The search for $R$-parity conserving supersymmetric particles at colliding beam
experiments is plagued by the necessity to {\it pair produce} sparticles,
and by the fact that the sparticle cascade decay terminates in the
lightest SUSY particle (LSP), usually assumed to comprise at least a portion of the
missing dark matter in the universe. The first of these thus requires enough
energy to produce two rather than just one sparticle, while the second of
these means that the sparticle invariant mass can't be directly
reconstructed as a resonance.
An alternative path to SUSY discovery at collider experiments is to search
for the $R$-parity even neutral heavy Higgs bosons, the heavy scalar $H$ and
the pseudoscalar $A$. These particles can be produced singly as $s$-channel
resonances and have the advantage in that their invariant mass can,
in principle, be directly reconstructed
(as was the case in discovery of the light scalar $h$).

In this paper, we examine production and decay of the heavy neutral
scalar Higgs bosons of the MSSM in the most lucrative discovery channel
$pp\to H,\ A\to\tau\bar{\tau}$. In previous phenomenological work\cite{Carena:2013qia,Djouadi:2013uqa,Djouadi:2015jea,Bagnaschi:2018ofa,Bahl:2020kwe},
new scenarios were proposed for the $m_A$ vs. $\tan\beta$ discovery plane
which ensured that $m_h\simeq 125$ GeV while also respecting that LHC
sparticle search limits were enforced, usually by assuming supersymmetry
breaking in the multi-TeV regime. These constraints can in principle
affect the regions of the heavy Higgs search planes which can be probed by
current and forthcoming hadron colliders.

In the present work, we add to these constraints the condition that the
magnitude of the weak scale also be {\it natural}.
This is because natural SUSY models are in a sense more plausible than
unnatural models\cite{Baer:2022dfc}.
For our naturalness criterion, we adopt the notion of practical naturalness\cite{Baer:2015rja}:
\begin{quotation}
  An observable ${\cal O}=o_1 +\cdots +o_n$ is natural if all {\it independent} contributions to ${\cal O}$ are comparable to or less than ${\cal O}$.
  \end{quotation}
Here, we adopt the measured value of the $Z$-boson mass as representative of
the magnitude of weak scale, where in the Minimal Supersymmetric Standard Model
(MSSM)\cite{Baer:2006rs}, the $Z$ mass is related to Lagrangian parameters
via the electroweak minimization condition
\be
m_Z^2/2 =\frac{m_{H_d}^2+\Sigma_d^d-(m_{H_u}^2+\Sigma_u^u )\tan^2\beta}{\tan^2\beta -1}-\mu^2
\label{eq:mzs}
\ee
where $m_{H_u}^2$ and $m_{H_d}^2$ are the Higgs soft breaking masses,
$\mu$ is the (SUSY preserving) superpotential $\mu$ parameter and
the $\Sigma_d^d$ and $\Sigma_u^u$ terms contain a large assortment of
loop corrections (see Appendices of Ref's \cite{Baer:2012cf} and \cite{Baer:2021tta}
and also \cite{Dedes:2002dy} for leading two-loop corrections).
For natural SUSY models, the naturalness measure\cite{Baer:2012up}
\be
\Delta_{EW}\equiv |maximal\ term\ on\ RHS\ of\ Eq.~\ref{eq:mzs}|/(m_Z^2/2)
\ee
is adopted here where a value
\be
\Delta_{EW}\alt 30
\label{eq:dew30}
\ee
fulfills the {\it comparable} condition of practical naturalness.
For most SUSY benchmark models, the superpotential $\mu$ parameter is tuned
to cancel against large contributions to the weak scale from SUSY breaking.
Since the $\mu$ parameter typically arises from very different physics
than SUSY breaking, {\it e.g.} from whatever solution to the SUSY
$\mu$ problem that is assumed,\footnote{Twenty solutions to the SUSY
  $\mu$ problem are recently reviewed in Ref. \cite{Bae:2019dgg}.}
then such a ``just-so'' cancellation
seems highly implausible\cite{Baer:2022dfc}
(though not impossible) compared to the
case where all contributions to the weak scale are $\sim m_{weak}$,
so that $\mu$ (or any other parameter) need not be tuned.

There are several important implications of Eq. \ref{eq:dew30} for
heavy neutral SUSY Higgs searches.
\begin{itemize}
\item The superpotential $\mu$ parameter enters $\Delta_{EW}$ directly,
  leading to $|\mu |\alt 350$ GeV.
  This implies that for heavy Higgs searches with $m_{A,H}\agt 2|\mu |$, then
  SUSY decay modes of $H,\ A$ should typically be open. If these additional
  decay widths to SUSY particles are large, then the branching fraction to the
  $\tau\bar{\tau}$ discovery mode can be substantially reduced.
\item For $m_{H_d}\gg m_{H_u}$, then $m_{H_d}$ sets the heavy Higgs mass scale
  ($m_{A,H}\sim m_{H_d}$) while $m_{H_u}$ sets the mass scale for $m_{W,Z,h}$.
  Then naturalness requires\cite{Bae:2014fsa}
    \end{itemize}
\be
m_{A,H}\alt m_Z\tan\beta\sqrt{\Delta_{EW}}.
\ee
For $\tan\beta\sim 10$ with $\Delta_{EW}\alt 30$, then $m_A$ can range up
to $\sim 5$ TeV. For $\tan\beta\sim 40$, then $m_A$ stays natural up to
$\sim 20$ TeV (although for large $\tan\beta\agt 20$, then bottom squark
contributions to $\Sigma_u^u$ become large and provide typically much
stronger limits on natural SUSY spectra).
Since most $H,A\to\tau\bar{\tau}$ searches and projected reach limits
take place assuming a decoupled SUSY spectra, then such results can
overestimate the collider heavy Higgs reach since in general the presence
of $H,A\to SUSY$ decay modes will diminish the $H,\ A\to\tau\bar{\tau}$
branching fraction.

Using naturalness, in Sec. \ref{sec:natplane} we propose a new natural SUSY
benchmark scenario $m_h^{125}({\rm nat})$ which is also consistent with
expectations from the string landscape\cite{Baer:2020kwz}.
In Sec. \ref{sec:prodBF}, we discuss production and decay of heavy neutral
Higgs bosons in the $m_h^{125}({\rm nat})$ scenario.
In Sec. \ref{sec:mT} we discuss signal event generation and SM backgrounds
for the case of back-to-back (BtB) $\tau$s in the transverse plane
using the total transverse mass variable $m_T^{tot}$.
In Sec. \ref{sec:mtautau}, we discuss signal and background for
the acollinear tau pairs using the $m_{\tau\tau}$ variable.
Including this signal channel can lead to a substantial increase in
signal significance and so combined with the BtB  $\tau$s can give an
increased collider reach in the $m_A$ vs. $\tan\beta$ search plane.
In Sec. \ref{sec:LHCreach}, we present our reach of present LHC
with 139 fb$^{-1}$ and also the projected reach of LHC Run3 and
HL-LHC.
%while in Sec. \ref{sec:FCCreach} we prsent reach projections for
%a future circular hadron collider (FCChh) with $\sqrt{s}=50$ or 100 TeV and 15 ab$^{-1}$.
Our conclusions reside in Sec. \ref{sec:conclude}.

\section{The natural SUSY Higgs search plane}
\label{sec:natplane}

The mass of the light SUSY Higgs boson is given approximately by\cite{Slavich:2020zjv}
\be
m_h^2\simeq m_Z^2\cos^22\beta +\frac{3g^2}{8\pi^2}\frac{m_t^4}{m_W^2}
\left[\ln\frac{m_{\tst}^2}{m_t^2}+\frac{x_t^2}{m_{\tst}^2}\left(1-\frac{x_t^2}{12m_{\tst}^2}\right)\right]
\ee
where $x_t=A_t-\mu\cot\beta$ and $m_{\tst}^2\simeq m_{Q_3}m_{U_3}$ is the mean top
squark mass. For a given value of $m_{\tst}^2$, then $m_h^2$ is maximal for
$x_t^{max}=\pm\sqrt{6}m_{\tst}$.

\subsection{Some previous SUSY Higgs benchmark studies}

In Ref. \cite{Carena:2013qia}, a variety of SUSY Higgs search benchmark
points were proposed, including 1. the $m_h^{max}$ scenario where a value of
$x_t^{max}$ was chosen along with $m_{\tg}=1500$ GeV and $m_{SUSY}\equiv m_{\tst}=1$ TeV with $\mu=M_2=0.2$ TeV as a conservative choice which maximized the
parameter space of the $m_A$ vs. $\tan\beta$ plane available for new
SUSY Higgs boson searches. Similarly, $m_h^{mod+}$ and $m_h^{mod-}$ scenarios
were proposed with similar parameters except for more moderate
$x_t=1.6 m_{SUSY}$ and $x_t=-2.2 m_{SUSY}$ values. Light stop, light stau,
$\tau$-phobic and low $m_H$ scenarios were proposed as well. Over time,
all these benchmark models have become LHC-excluded since (at least)
they all proposed $m_{\tg}\sim 1500$ GeV while after LHC Run 2 the ATLAS/CMS
Collaborations require $m_{\tg}\agt 2.2$ TeV\cite{ATLAS:2020syg,CMS:2019zmd}.

In Ref. \cite{Bagnaschi:2018ofa}, an $m_h^{125}$ benchmark model was proposed
with $m_{SUSY}\sim 1.5$ TeV, $\mu =1$ TeV and $m_{\tg}=2.5$ TeV in accord
with LHC Run 2 gluino mass constraints.
The $x_t=2.8$ TeV value was chosen to nearly maximize the value of $m_h$
given the other parameters of the model. This model has almost all
$H,\ A\to SUSY$ decay modes kinematically closed due to the heavy SUSY spectra
so it closely resembles the type-II two-Higgs doublet model (2HDM)  phenomenology\cite{Branco:2011iw}.
An $m_h^{125}(\tilde{\tau} )$ scenario
(exemplifying bino-stau coannihilation was selected with $\mu=1$ TeV
along with a $m_h^{125}(\tchi )$ scenario with $\mu =180$ GeV, $M_1=160$ GeV
and $M_2=180$ GeV so that $H,\ A$ decay to many electroweakino states is allowed. Also, an $m_h^{125}(align)$ model with specific {\it alignment without decoupling}\cite{Gunion:2002zf,Carena:2013ooa} parameters with $\mu =7.5$ TeV was chosen along with a
$m_H^{125}$  scenario where the heavy Higgs scalar was actually the 125 GeV Higgs boson. These scenarios would be hard pressed to explain why
$m_{weak}\sim 100$ GeV due to the tuning needed for such large
$\mu$ parameters. The exception is the $m_h^{125}(\tchi )$ scenario,
although here the peculiar gaugino/higgsino mass choices seem at odds with
most theory expectations\footnote{Gaugino mass unification is usually expected
  in models based on grand unification, but is also expected by the simple
  form of the supergravity (SUGRA) gauge kinetic function which depends typically on only a single hidden sector field in many string-inspired constructs.}.

A somewhat different approach is taken in the model labelled $hMSSM$\cite{Djouadi:2013uqa,Djouadi:2015jea,Arcadi:2022hve}.
In the hMSSM, by adopting a high $m_{SUSY}$ scale and by neglecting
some small radiative corrections to the
Higgs mass matrix, then one may use $m_h$ (along with $m_A$ and $\tan\beta$)
as an input parameter with Higgs mixing angle $\alpha$, $m_H$
and $m_{H^{\pm}}$ as outputs.
This ensures that $m_h=125$ GeV is enforced throughout the remaining
Higgs search parameter space. The adoption of a high value $m_{SUSY}\agt 1$ TeV
then makes this model look like the 2HDM, and sparticle mass spectra are
effectively neglected. By combining $H,\ A\to\tau\bar{\tau}$ with
$H,\ A\to t\bar{t}$ at lower $\tan\beta$, then it is claimed almost the entire
$m_A$ vs. $\tan\beta$ parameter space can be probed by HL-LHC for
$m_A\alt 1$ TeV\cite{Djouadi:2015jea}.

\subsection{Status of Run 2 LHC searches}

The ATLAS Collaboration has reported on a search for
$H,\ A\to\tau\bar{\tau}$ at CERN LHC Run 2 using 139 fb$^{-1}$ of
integrated luminosity at $\sqrt{s}=13$ TeV\cite{ATLAS:2020zms}.
The study focusses on
back-to-back $\tau\bar{\tau}$ states where transverse opening angles
$\Delta\phi (\tau_{had}\tau_{had})>155^\circ$ and $\Delta\phi (\tau_{lep}\tau_{had})>135^\circ$ are required. Mixed leptonic-hadronic
($\tau_{lep}\tau_{had}$) and hadronic-hadronic ($\tau_{had}\tau_{had}$) final
states are combined.
The hadronic tau tagging efficiency in one or three charged prong
$\tau$-jets varies from 60-85\%.
The total transverse mass\cite{Barger:1984sm}
\be
m_T^{tot}=\sqrt{(p_T^{\tau_1}+p_T^{\tau_2}+E_T^{miss})^2-(\vec{p}_T^{\tau_1}+\vec{p}_T^{\tau_2}+\vec{E}_T^{miss})^2}
\ee
is measured and a fit to expected signal plus background is made to determine
the presence of a signal. For the signal, the $m_T^{tot}$ distribution is
bounded from above by $m_T^{tot}< m_{H,\ A}$ and near this upper bound is where
the signal-to-background significance is greatest. In this region,
the dominant background comes from Drell-Yan $\gamma^*,\ Z\to\tau\bar{\tau}$
production.
The signal sample is further divided by either the presence or absence of
a tagged $b$-jet but the signal significance is dominated by the $b$-jet vetoed
events. No signal is found, so the 95\% CL
exclusion limits are plotted in the $m_A$ vs. $\tan\beta$ plane in the
Bagnaschi {\it et al.} $m_h^{125}$ scenario\cite{Bagnaschi:2018ofa}.
They find that for $\tan\beta\sim 10$, then $m_A\alt 1.1$ TeV is already
excluded while for $\tan\beta\sim 50$, then $m_A\alt 2$ TeV is excluded.

The CMS collaboration has presented results of $H,\ A\to\tau\bar{\tau}$
searches using 35.9 fb$^{-1}$ of integrated luminosity\cite{CMS:2018rmh}.
The 95\% CL exclusion limits are plotted in the $m_A$ vs. $\tan\beta$
plane for the $m_h^{mod+}$ and hMSSM scenarios.
Further CMS analyses using the full Run 2 data set should be forthcoming.

\subsection{Some previous LHC upgrade SUSY Higgs reach studies}

In Ref. \cite{Cepeda:2019klc}, 
%{\it Report from Working Group 2: Higgs Physics at the HL-LHC and HE-LHC}, 
the ATLAS and CMS collaborations
presented expected reach plots for $H,\ A\to\tau\bar{\tau}$ for HL-LHC
with either 3 or 6 ab$^{-1}$ of integrated luminosity and $\sqrt{s}=14$ TeV.
The results were a direct extrapolation of their previous search results
from LHC Run 2. ATLAS with 3 ab$^{-1}$ expects to explore $m_A\alt 1500$ GeV
for $\tan\beta =10$ in the hMSSM scenario and up to $m_A\alt 1$ TeV in the
$m_h^{mod+}$ scenario. The plot upper limit of $m_A<2250$ GeV precludes
any limits for $\tan\beta \agt 40$.
With 3 ab$^{-1}$, the CMS collaboration expects to explore at 95\% CL up to
$m_A<750$ GeV in the $m_h^{mod+}$ scenario and up to $m_A\alt 1400$ GeV in the
hMSSM scenario, both for $\tan\beta =10$.

The HL-LHC and ILC sensitivity for heavy SUSY Higgs bosons was also
estimated by Bahl {\it et al.}\cite{Bahl:2020kwe}. Their 95\% CL
exclusion using a combined ATLAS/CMS sensitivity (6 ab$^{-1}$) is
to explore up to $m_A\alt 1500$ GeV for $\tan\beta =10$ in the $m_h^{125}$
scenario (heavy SUSY) and to $m_A\alt 1$ TeV in the $m_h^{125} (\tchi )$
scenario (light EWinos).
They also explore some $m_{h,EFT}^{125}$ scenarios\cite{Bahl:2019ago} 
at low $\tan\beta\sim 1-10$ which we will not consider.

\subsection{The $m_h^{125}(nat)$ Higgs search benchmark}

In this Subsection, we introduce a more plausible SUSY Higgs search
benchmark model in that all its contributions to the weak scale
are comparable to or less than the weak scale by a conservative factor
of $\sim 4$. This would be the class of natural SUSY models characterized by
$\Delta_{EW}\alt 30$\cite{Baer:2012up}. These natural SUSY models can be found in several
different guises:
\begin{enumerate}
\item The 2,3,4-extra parameter non-universal Higgs models NUHM2,3,4
  which characterize what might be expected from dominant gravity-mediated
  SUSY breaking\cite{Baer:2012cf},
\item  natural anomaly-mediated SUSY breaking\cite{Baer:2018hwa} (nAMSB) wherein non-universal bulk soft terms
  allow for naturalness while maintaining $m_h\simeq 125$ GeV and
  \item natural generalized
mirage-mediation (nGMM) models\cite{Baer:2016hfa} wherein soft terms are characterized by comparable
anomaly- and gravity/moduli-mediated contributions.
The nGMM model is expected to emerge\cite{Choi:2005ge} from KKLT moduli stabilization\cite{Kachru:2003aw} and the string landscape\cite{Broeckel:2020fdz}.
\end{enumerate}
For our benchmark models, it is perhaps easiest to settle on the more familiar
gravity-mediated two-extra-parameter non-universal Higgs model 
NUHM2\cite{Ellis:2002wv,Baer:2005bu} which is characterized by the parameter space
\be
m_0,\ m_{1/2},\ A_0,\ \tan\beta,\ m_{H_u},\ m_{H_d}
\ee
where $m_0$ denotes the GUT scale matter scalar soft terms,
$m_{1/2}$ are the unified gaugino masses, $A_0$ are common trilinear soft terms
and $\tan\beta\equiv v_u/v_d$ is the usual ratio of Higgs field vevs.
It is reasonable to have $m_{H_u}\ne m_{H_d}\ne m_0$ in gravity-mediation
since the scalar mass soft terms in supergravity do not respect universality.
However, a remnant $SO(10)$ local GUT symmetry may enforce the matter scalars of
each generation to have a common mass $m_0(i)$, where $i=1-3$ is a
generation index.\footnote{
  In the landscape context, the first two generations are pulled to common
  upper bounds which yields a mixed decoupling/quasi-degeneracy solution to the SUSY flavor and CP problems\cite{Baer:2019zfl}. The third generation is pulled up much less
  than the first two generations since it contributes more to the weak scale
  via the large Yukawa couplings.}
The Higgs soft terms $m_{H_u}$ and $m_{H_d}$ are frequently traded for the weak scale parameters $\mu$ and $m_A$ via the scalar potential minimization conditions.
Thus, the parameter space of NUHM2
\be
m_0,\ m_{1/2},\ A_0,\ \tan\beta,\ \mu,\ m_A
\ee
is well-suited to Higgs searches since it allows for variable 
$m_A$ and $\tan\beta$
as independent input parameters while also allowing the input of $\mu\alt 350$
GeV which is required by naturalness in Eq. \ref{eq:mzs}. 

Using NUHM2, we adopt the following natural SUSY benchmark Higgs search scenario:
\be
m_h^{125}({\rm nat}):\ m_0=5\ {\rm TeV},\ m_{1/2}=1.2\ {\rm TeV},\ A_0=-1.6m_0,\ \tan\beta ,\ \mu=250\ {\rm GeV}\ {\rm and}\ m_A .
\ee
The $m_h^{125}({\rm nat})$ benchmark model spectra is shown in Table \ref{tab:bm} for $\tan\beta =10$ and $m_A=2$ TeV.
We adopt the computer code Isajet\cite{Paige:2003mg} featuring Isasugra\cite{Baer:1994nc}
for spectra generation. The SUSY Higgs boson masses are computed
using renormalization-group (RG) improved third generation fermion/sfermion
loop corrections\cite{Bisset:1995dc}.
The RG improved Yukawa couplings include full
threshold corrections\cite{Pierce:1996zz} which account for leading
two-loop effects\cite{Carena:2002es}.
From the Table, we note that $m_h=124.7$ GeV and $\Delta_{EW}=22$.
Recent versions of FeynHiggs\cite{Bahl:2018qog} predict $m_h$ values closer to Isasugra
than past versions, and for the $m_h^{125}({\rm nat})$ benchmark
point we find from FeynHiggs 2.18.1 that $m_h=125.3\pm 1.3$ GeV,
in close accord with Isasugra.
%BM point : \TABLE{
\begin{table}[h!]
\centering
\begin{tabular}{lc}
\hline
parameter & $m_h^{125}({\rm nat})$ \\
\hline
$m_0$      & 5 TeV \\
$m_{1/2}$      & 1.2 TeV \\
$A_0$      & -8 TeV \\
$\tan\beta$    & 10  \\
\hline
$\mu$          & 250 GeV  \\
$m_A$          & 2 TeV \\
\hline
$m_{\tilde{g}}$   & 2830 GeV \\
$m_{\tilde{u}_L}$ & 5440 GeV \\
$m_{\tilde{u}_R}$ & 5561 GeV \\
$m_{\tilde{e}_R}$ & 4822 GeV \\
$m_{\tilde{t}_1}$& 1714 GeV \\
$m_{\tilde{t}_2}$& 3915 GeV \\
$m_{\tilde{b}_1}$ & 3949 GeV \\
$m_{\tilde{b}_2}$ & 5287 GeV \\
$m_{\tilde{\tau}_1}$ & 4746 GeV \\
$m_{\tilde{\tau}_2}$ & 5110 GeV \\
$m_{\tilde{\nu}_{\tau}}$ & 5107 GeV \\
$m_{\tilde{\chi}_1^\pm}$ & 261.7 GeV \\
$m_{\tilde{\chi}_2^\pm}$ & 1020.6 GeV \\
$m_{\tilde{\chi}_1^0}$ & 248.1 GeV \\ 
$m_{\tilde{\chi}_2^0}$ & 259.2 GeV \\ 
$m_{\tilde{\chi}_3^0}$ & 541.0 GeV \\ 
$m_{\tilde{\chi}_4^0}$ & 1033.9 GeV \\ 
$m_h$       & 124.7 GeV \\ 
\hline
$\Omega_{\tilde{z}_1}^{std}h^2$ & 0.016 \\
$BF(b\to s\gamma)\times 10^4$ & $3.1$ \\
$BF(B_s\to \mu^+\mu^-)\times 10^9$ & $3.8$ \\
$\sigma^{SI}(\tilde{\chi}_1^0, p)$ (pb) & $2.2\times 10^{-9}$ \\
$\sigma^{SD}(\tilde{\chi}_1^0, p)$ (pb)  & $2.9\times 10^{-5}$ \\
$\langle\sigma v\rangle |_{v\to 0}$  (cm$^3$/sec)  & $1.3\times 10^{-25}$ \\
$\Delta_{\rm EW}$ & 22 \\
\hline
\end{tabular}
\caption{Input parameters (TeV) and masses (GeV)
for the $m_h^{125}({\rm nat})$ SUSY benchmark point from the NUHM2 model
with $m_t=173.2$ GeV using Isajet 7.88~\cite{Paige:2003mg}.
}
\label{tab:bm}
\end{table}

\begin{figure}[htb!]
\begin{center}
  \includegraphics[height=0.25\textheight]{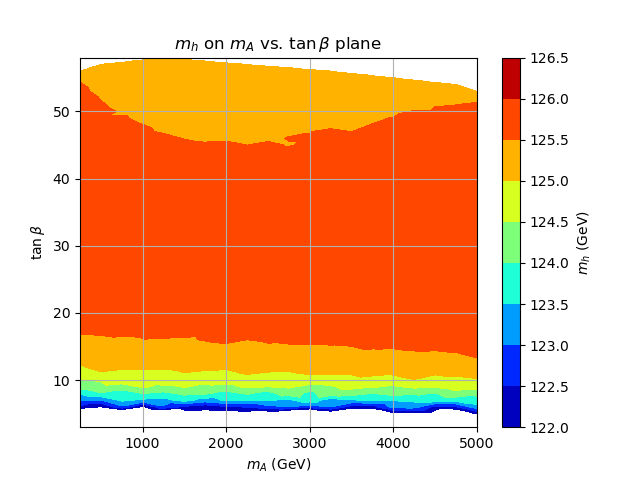}
  \includegraphics[height=0.25\textheight]{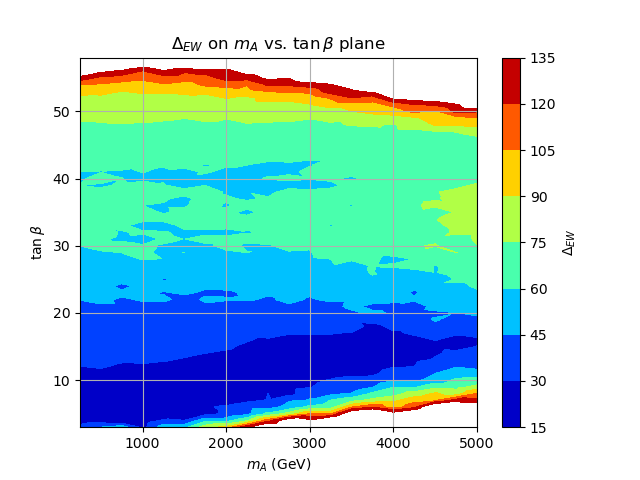}
\caption{{\it a}) Contours of $m_h$ in the $m_A$ vs. $\tan\beta$ plane using the
  $m_h^{125}({\rm nat})$ scenario from the NUHM2 model with $m_0=5$ TeV, $m_{1/2}=1.2$ TeV,
  $A_0=-8$ TeV and $\mu=250$ GeV. {\it b}) Regions of electroweak naturalness
  measure $\Delta_{EW}$ in the same plane as {\it a}).
  \label{fig:mhdew}}
\end{center}
\end{figure}

In Fig. \ref{fig:mhdew}{\it a}), we show regions of light Higgs mass
$m_h$ in the $m_A$ vs. $\tan\beta$ plane for the $m_h^{125}({\rm nat})$
benchmark scenario. From the plot, we can see that the value of $m_h$
is indeed very close to 125 GeV throughout the entire plane except for
very low $\tan\beta\alt 6$ where $m_h$ dips below $123$ GeV.
In Fig. \ref{fig:mhdew}{\it b}), we show regions of EW naturalness
measure $\Delta_{EW}$. We see that in the region of $\tan\beta:1-15$,
then $\Delta_{EW}\alt 30$ even for $m_A$ extending out as high as
$5$ TeV. For larger $\tan\beta\agt 20$, then $\Delta_{EW}$ moves to
$\sim 45-90$ mainly because the $b$- and $\tau$-Yukawa couplings grow
and lead to large $\Sigma_u^u(\tb,\ttau )$ terms.

\section{Production and decay of $H,\ A$ in the $m_h^{125}({\rm nat})$ scenario}
\label{sec:prodBF}

\subsection{$H$ and $A$ production cross sections in the $m_h^{125}({\rm nat})$ scenario}

\begin{figure}[htb!]
\begin{center}
\includegraphics[height=0.25\textheight]{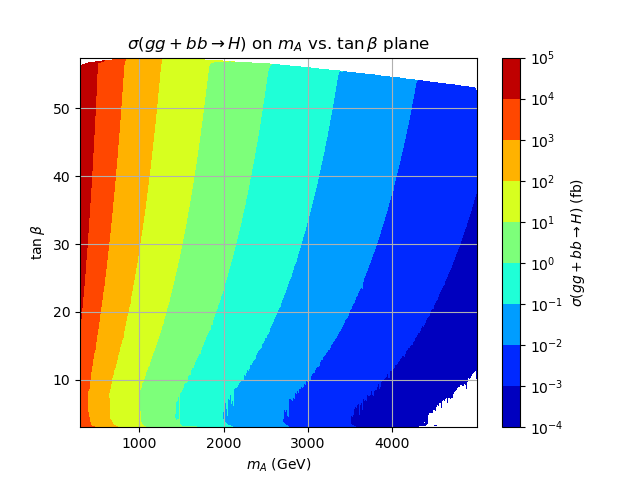}
\includegraphics[height=0.25\textheight]{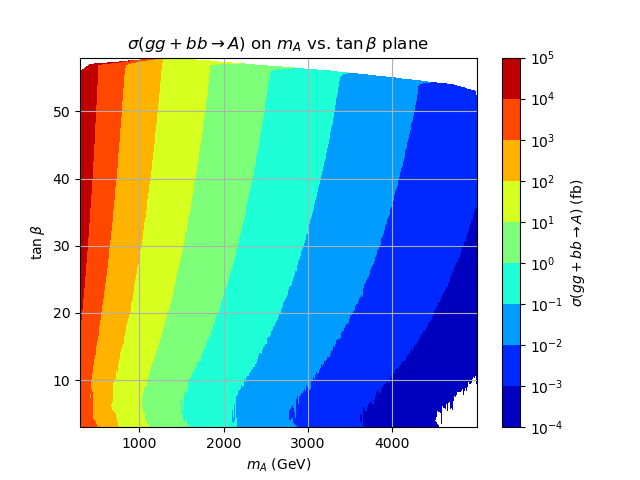}\\
\caption{The total cross section for {\it a}) $pp\to H$
  and {\it b}) $pp\to A$ at $\sqrt{s}=14$ TeV using the SusHi
  code\cite{Harlander:2012pb}.
\label{fig:sigHA14}}
\end{center}
\end{figure}

The $s$-channel resonance production of the $H$ and $A$ bosons takes place
mainly via $gg$ and $q\bar{q}$ (mainly $b\bar{b}$) fusion reactions at
hadron colliders. The total $H$ and $A$ production cross sections
are shown in the $m_A$ vs. $\tan\beta$ plane in Fig. \ref{fig:sigHA14}
for $\sqrt{s}=14$ TeV $pp$ collisions-- as are expected at CERN LHC Run 3
and at high-luminosity LHC (HL-LHC) where of order 300 fb$^{-1}$ (for Run 3)
and 3000 fb $^{-1}$ (for HL-LHC) of integrated luminosity is expected
to be obtained. 
For the cross sections, we use the computer code
SusHi\cite{Harlander:2012pb} which contains contributions up to
NNLO in perturbative QCD. The cross sections range from over
$10^4$ fb at low $m_{H,A}\sim 400$ GeV down to $\sigma (pp\to H,A )<1$ fb for
$m_{H,A}\sim 2$ TeV, and they increase somewhat with increasing $\tan\beta$
where production via $b\bar{b}$ fusion is enhanced.

\subsection{$H$ and $A$ branching fractions in the $m_h^{125}({\rm nat})$ scenario}

It is sometimes claimed in the literature that the tree-level
production and decay rates for the $H$ and $A$ bosons depend only
on $m_A$ and $\tan\beta$, and indeed search limits for the heavy Higgs
bosons are typically presented in the $m_A$ vs. $\tan\beta$
plane, following the early pioneering work by Kunszt and
Zwirner\cite{Kunszt:1991qe}.
While this is true for the (non-supersymmetric) 2HDM, it is not true for the
MSSM, where the importance of tree level SUSY Higgs boson decays to SUSY
particles was first emphasized in \cite{Baer:1987eb,Gunion:1987ki,Gunion:1988yc}. In the 2HDM, decays of $H$ and $A$ to the heaviest available
fermion pairs will typically dominate, with decays to $b\bar{b}$
and $\tau\bar{\tau}$ enhanced at large $\tan\beta$. However, in SUSY models
there is a direct gauge coupling
\be
{\cal L}\ni-\sqrt{2}\sum_{i,A}{\cal S}_i^\dagger g t_A\bar{\lambda}_A\frac{}{}\psi_i +H.c.
\ee
where ${\cal S}_i$ labels various matter and Higgs scalar fields
(labelled by $i$), $\psi_i$ is the fermionic superpartner of ${\cal S}_i$
and $\lambda_A$ is the gaugino with gauge index $A$.
Also, $g$ is the corresponding gauge coupling for the gauge group in
question and the $t_A$ are the corresponding gauge group matrices.
Letting ${\cal S}_i$ be the Higgs scalar fields,
we see there is an unsuppressed coupling of the Higgs scalars to
gaugino plus higgsino.
This coupling can lead to dominant SUSY Higgs boson decays to SUSY
particles when the gaugino-plus-higgsino decay channel is
kinematically allowed.

\begin{figure}[htb!]
\begin{center}
\includegraphics[height=0.25\textheight]{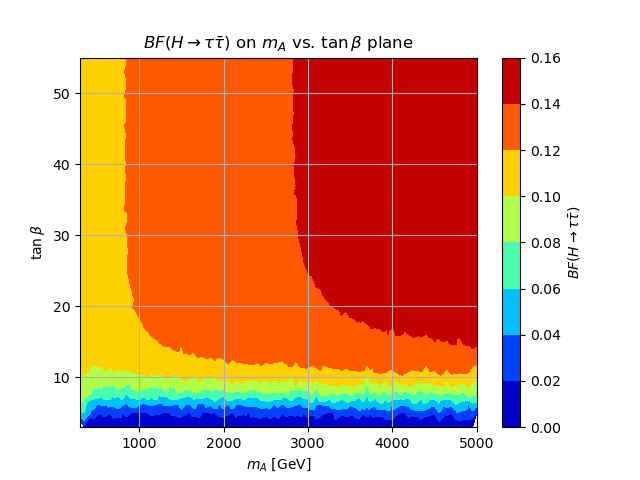}
\includegraphics[height=0.25\textheight]{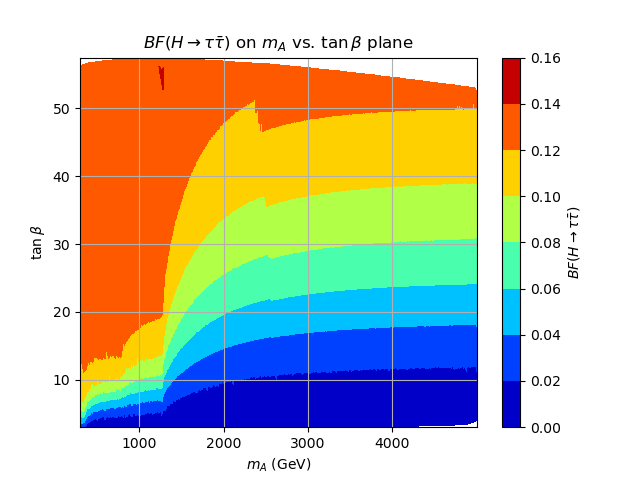}\\
\caption{Branching fraction of $H\to\tau\bar{\tau}$
  in the {\it a}) hMSSM and {\it b}) in the $m_h^{125}({\rm nat})$ benchmark case
  in the $m_A$ vs. $\tan\beta$ plane.
\label{fig:BFHtt1}}
\end{center}
\end{figure}

In Fig. \ref{fig:BFHtt1}, we plot the $H \to\tau\bar{\tau}$ branching
fractions as color-coded regions in the
$m_A$ vs. $\tan\beta$ plane for {\it a}) the hMSSM and {\it b}) for
our $m_h^{125}({\rm nat})$ BM scenario.
For the hMSSM, we use the computer code 2HDMC\cite{Eriksson:2009ws} 
with $m_h=125$ GeV throughout the $m_A$ vs. $\tan\beta$ plane but with decoupled sparticles. We use the ``Physical mass input set''. With the potential parameters $\lambda_i$ as in the tree-level MSSM except $\lambda_2$, which includes a correction term to bring the light CP-even higgs mass to be 125 GeV, the only free physical inputs left are then just $m_A$ and $\tan\beta$.
From frame {\it a}), we see
as expected that for the hMSSM, the $BF(H\to\tau\bar{\tau} )$ increases
with $\tan\beta$. It also increases slightly as $m_A$ increases
since the $\tau$ Yukawa coupling $f_\tau$ increases slightly with scale choice.
In frame {\it b}) for the $m_h^{125}({\rm nat})$ case, we again see an increasing branching fraction as $\tan\beta$ increases, but now as $m_A$
(and hence $m_H$) increases, various SUSY decay modes to EWinos open
up, especially around $m_A\sim 1200$ GeV where decays to gaugino-plus-higgsino
become accessible. We see the $BF(H\to\tau\bar{\tau})$ can drop from 12\%
on the left-side of the plot down to just a few percent on the right-hand-side.
This is due to the fact that the decay to EWinos ultimately dominates
the heavy Higgs branching fraction\cite{Bae:2015nva,Baer:2021qxa}.
There is also a glitch apparent at around $m_A\sim 2500$ GeV in the contours.
This occurs because we include SUSY threshold corrections to the
Yukawa couplings which are implemented at the scale
$m_{SUSY}^2=m_{\tst_1}m_{\tst_2}$ and so the Yukawa couplings have a
slight discontinuity (see {\it e.g.} Fig. 6 of Ref. \cite{Baer:2012jp}).

\begin{figure}[htb!]
\begin{center}
\includegraphics[height=0.25\textheight]{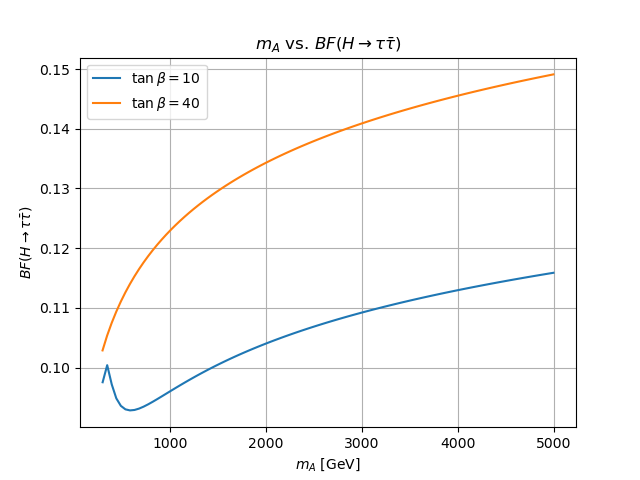}
\includegraphics[height=0.25\textheight]{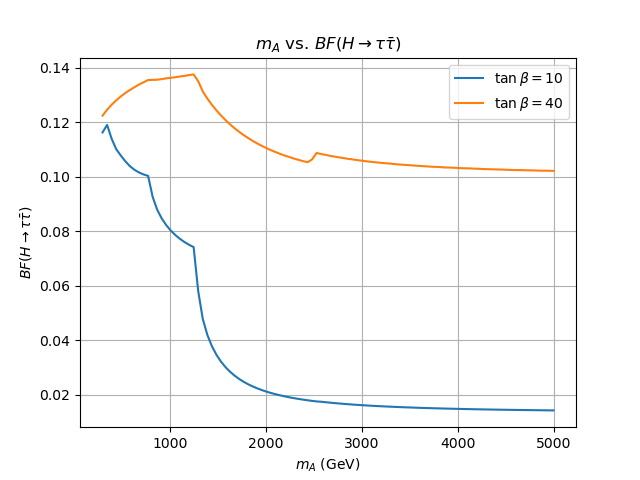}\\
\caption{Branching fraction of $H\to\tau\bar{\tau}$
  in the {\it a}) hMSSM and {\it b}) in the $m_h^{125}({\rm nat})$  benchmark case
  vs. $m_A$ for $\tan\beta =10$ and 40.
\label{fig:BFHtt2}}
\end{center}
\end{figure}

It is also helpful to show the explicit $BF(H\to\tau\bar{\tau})$ vs. $m_A$
for two specific choices of $\tan\beta =10$ and 40 for the {\it a})
hMSSM and {\it b}) the $m_h^{125}({\rm nat} )$ model in Fig. \ref{fig:BFHtt2}.
For the hMSSM, we again see the slight increase with increasing $m_A$,
although the BFs stay in the vicinity of 10-15\%.
For the $m_h^{125}({\rm nat})$ case, we see the sharp drop off in
$BF(H\to\tau\bar{\tau})$ as various $H\to EWinos$ thresholds are passed:
then, ultimately the branching fraction drops below 2\% for
large $m_A$.

\begin{figure}[htb!]
\begin{center}
\includegraphics[height=0.25\textheight]{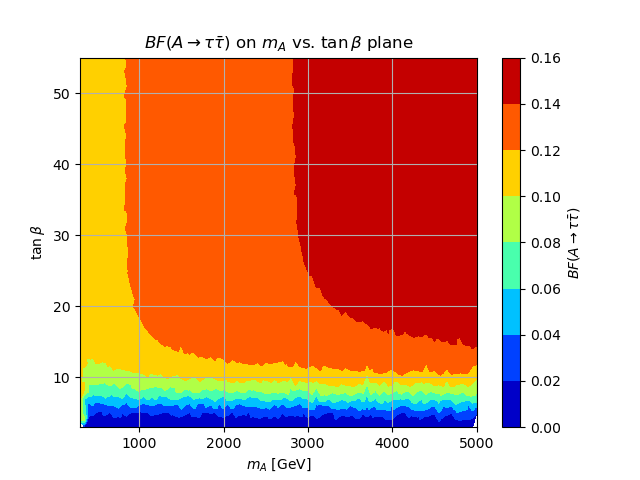}
\includegraphics[height=0.25\textheight]{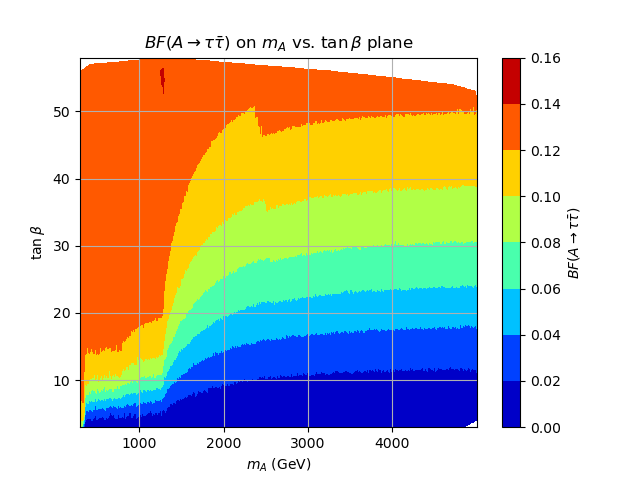}\\
\caption{Branching fraction of $A\to\tau\bar{\tau}$
  in the {\it a}) hMSSM and {\it b}) in the $m_h^{125}({\rm nat})$  benchmark case
  in the $m_A$ vs. $\tan\beta$ plane.
\label{fig:BFAtt1}}
\end{center}
\end{figure}

Similar behavior is shown in Fig. \ref{fig:BFAtt1}{\it a}) and {\it b})
for the $A\to\tau\bar{\tau}$ branching fraction: it has a slight increase with increasing $m_A$ for the hMSSM case but suffers sharp drops in the
$m_h^{125}({\rm nat})$ case due to the turn on of $A$ decay to
gaugino-plus-higgsino. This will affect the reach plots in a substantial way.

\begin{figure}[htb!]
\begin{center}
\includegraphics[height=0.25\textheight]{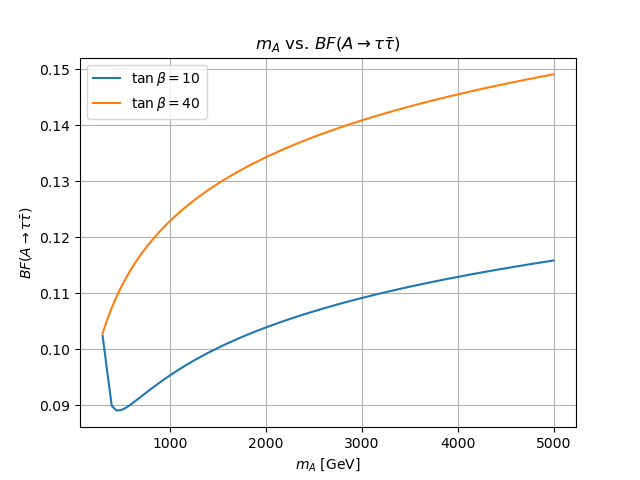}
\includegraphics[height=0.25\textheight]{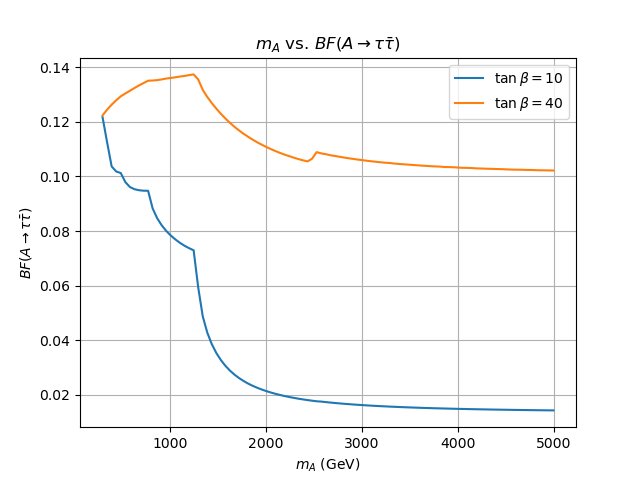}\\
\caption{Branching fraction of $A\to\tau\bar{\tau}$
  in the {\it a}) hMSSM and {\it b}) in the $m_h^{125}({\rm nat})$  benchmark case
  vs. $m_A$ for $\tan\beta =10$ and 40.
\label{fig:BFAtt2}}
\end{center}
\end{figure}

The corresponding plots of $BF(A\to\tau\bar{\tau})$ vs. $m_A$ for
$\tan\beta =10$ and 40 are shown in Fig. \ref{fig:BFAtt2}.
The behavior is rather similar to that already explained for the
$H$ decay.

\section{Signal from back-to-back $\tau\bar{\tau}$ via $m_T$}
\label{sec:mT}

In this Section, we present details from our event generation calculations
for the $H,\ A\to\tau\bar{\tau}$ signal with nearly back-to-back
(BtB) $\tau$s.
For signal and background event generation, we adopt the Pythia 8.07 event
generator\cite{Sjostrand:2007gs} interfaced with the Delphes toy detector simulation\cite{deFavereau:2013fsa}.
For signal, we generate $pp\to H,\ A\to\tau\bar{\tau}$
events with the total cross section adjusted to the SusHi NNLO result.
For SM backgrounds, we generate $q\bar{q}\to \gamma^*,Z\to\tau\bar{\tau}$
(Drell-Yan), $t\bar{t}$ and $VV$ production where $VV=W^+W^-,W^\pm Z$ and $ZZ$.

For jet finding, we use the Delphes FASTJET jet finder.
The FASTJET jet finder requires $p_T(jet)>25$ GeV and $\Delta R$ between jets
as $\Delta R_{jj}>0.4$. We also require $|\eta_{jet}|<2.5$.
Delphes includes a hadronic $\tau$-jet finding tool which we also use
which identifies one-and-three charged prong jets as tau jets provided
the tau is within $\Delta R=0.4$ of the jet in question. The Delphes
$\tau$-jet identification efficiency is found to be in the 50\% range
which is well below the ATLAS quoted $\tau$-jet efficiency ID
which is at the 75\% level. We also use the Delphes b-tag algorithm and the Delphes isolated lepton tag which requires
$\Delta R (l, l) > 0.3$ with $|\eta (e,\mu )|<2.5$. 

The $\tau_{had}\tau_{had}$ channel are selected by single-$\tau$ trigger $p_T$ cut of 160 GeV. Events contain at least two $\tau_{had}$ identified by the Delphes tau-tag algorithm. The two tau $\tau_{had}$ candidates must have opposite electric charge.

The $\tau_{lep}\tau_{had}$ channel are selected using single-electron and single-muon triggers with $p_T$ threshold of 30 GeV. The events contain exactly one isolated lepton and at least one $\tau_{had}$ candidate. The isolated lepton and the $\tau_{had}$ candidate must have opposite electric charge. Also, we rejected the events that the isolated lepton and the $\tau_{had}$ candidate have an invariant mass between 80 GeV and 110 GeV to reduce the background contribution from $Z\rightarrow ee$.

The events from either channel are further divided into categories of the $b$-tag for events containing at least one $b$-jet and the $b$-veto for events containing no $b$-jets.

\begin{figure}[htb!]
\begin{center}
\includegraphics[height=0.25\textheight]{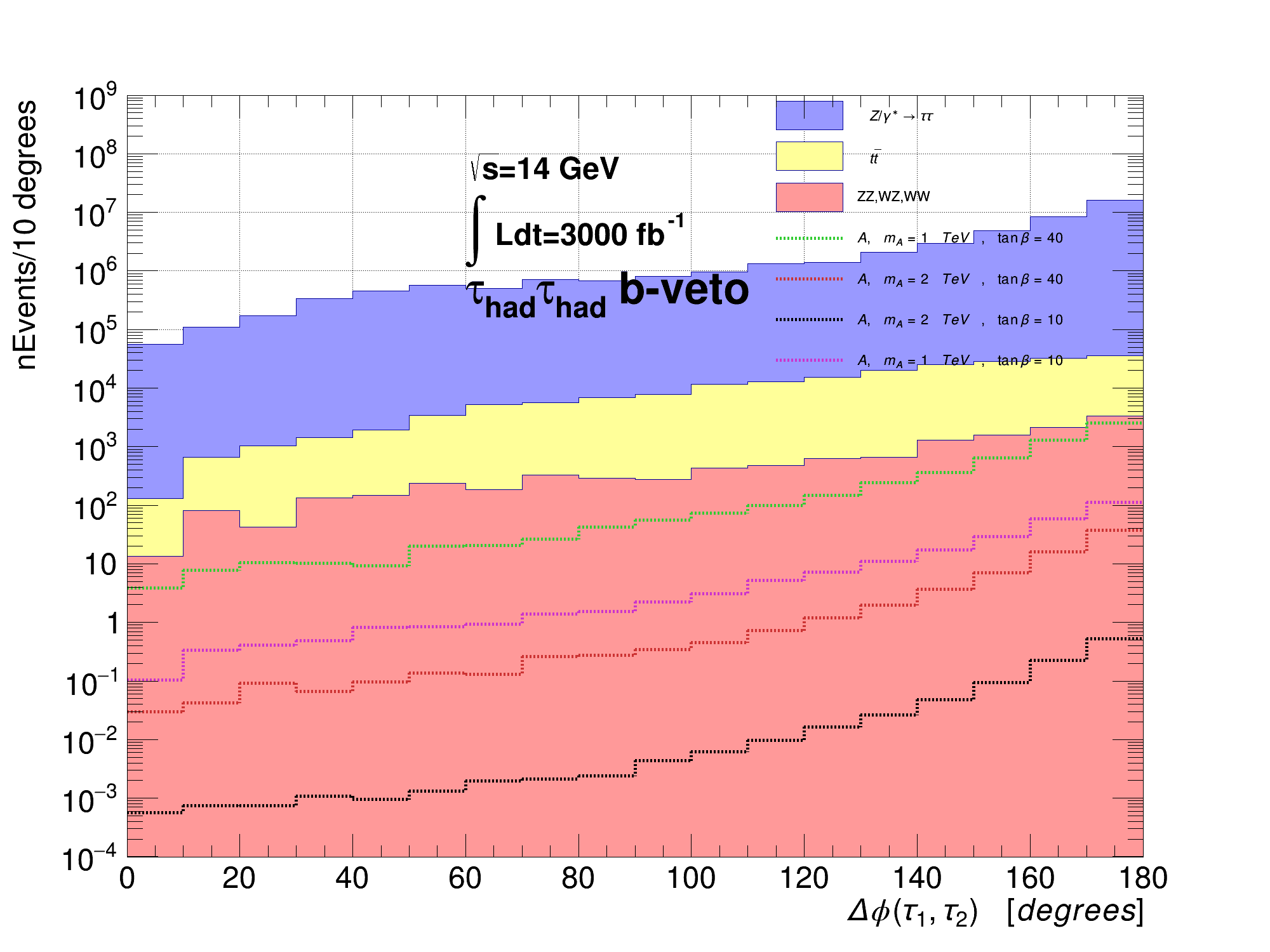}
\includegraphics[height=0.25\textheight]{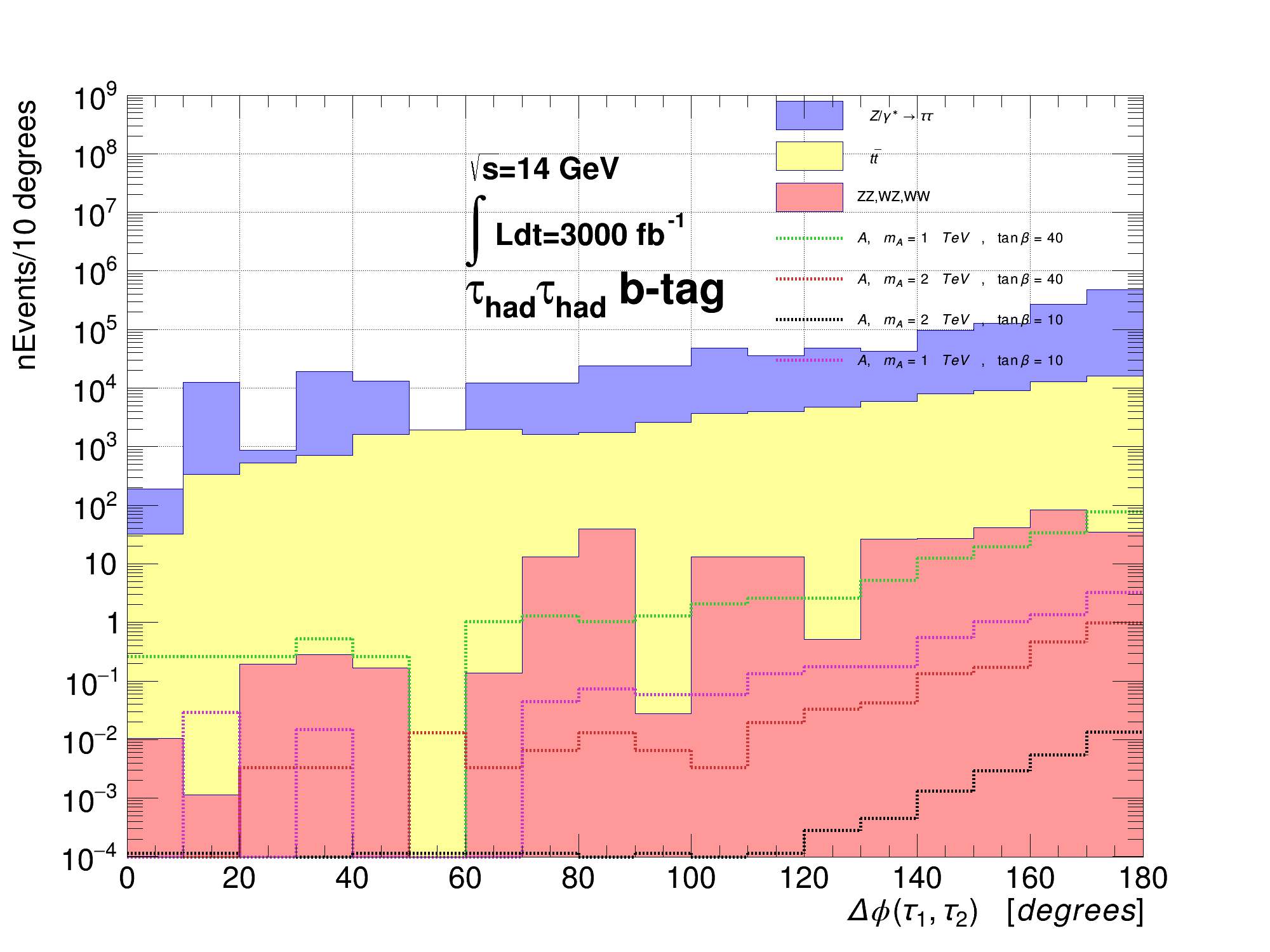}\\
\includegraphics[height=0.25\textheight]{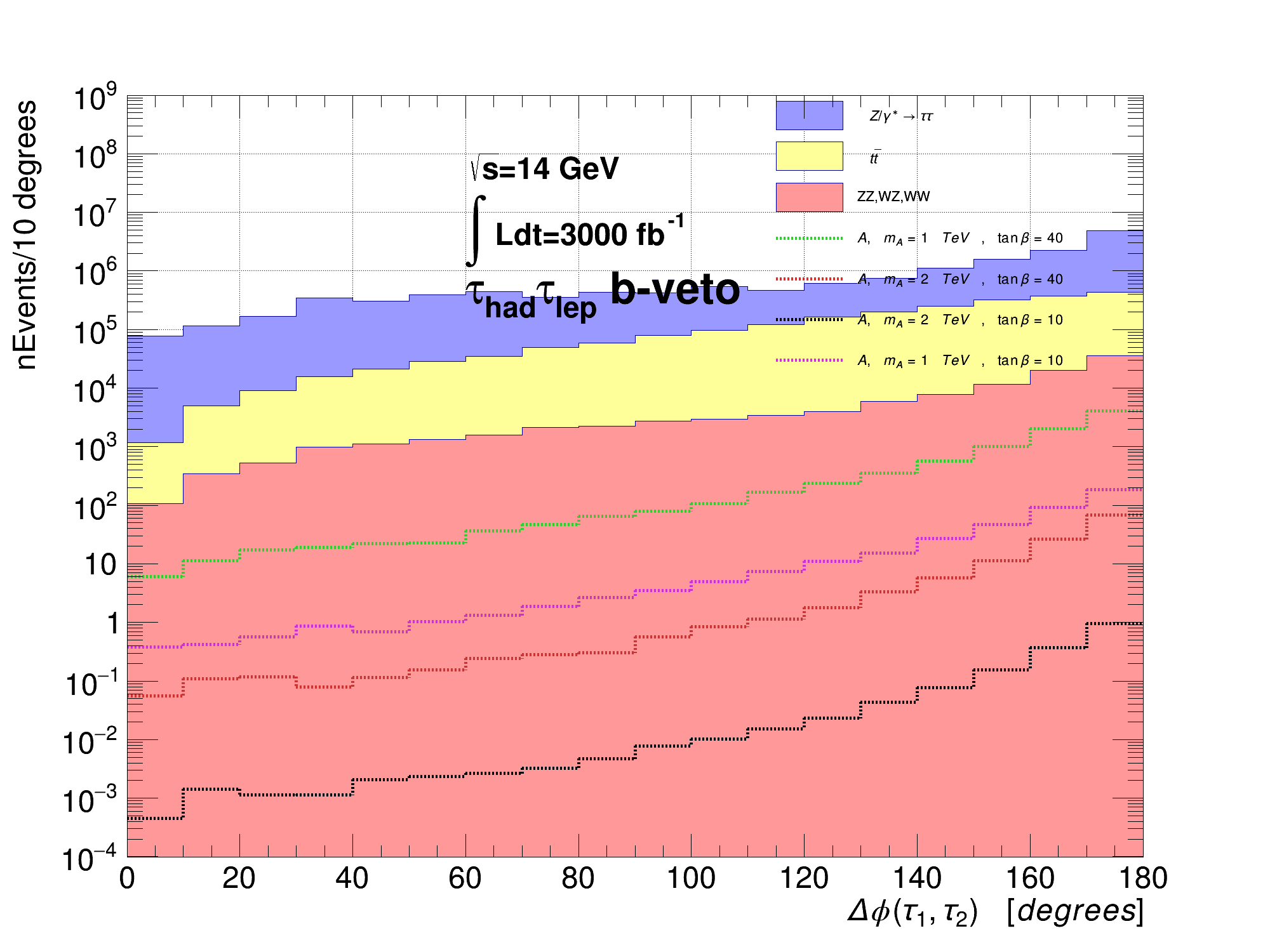}
\includegraphics[height=0.25\textheight]{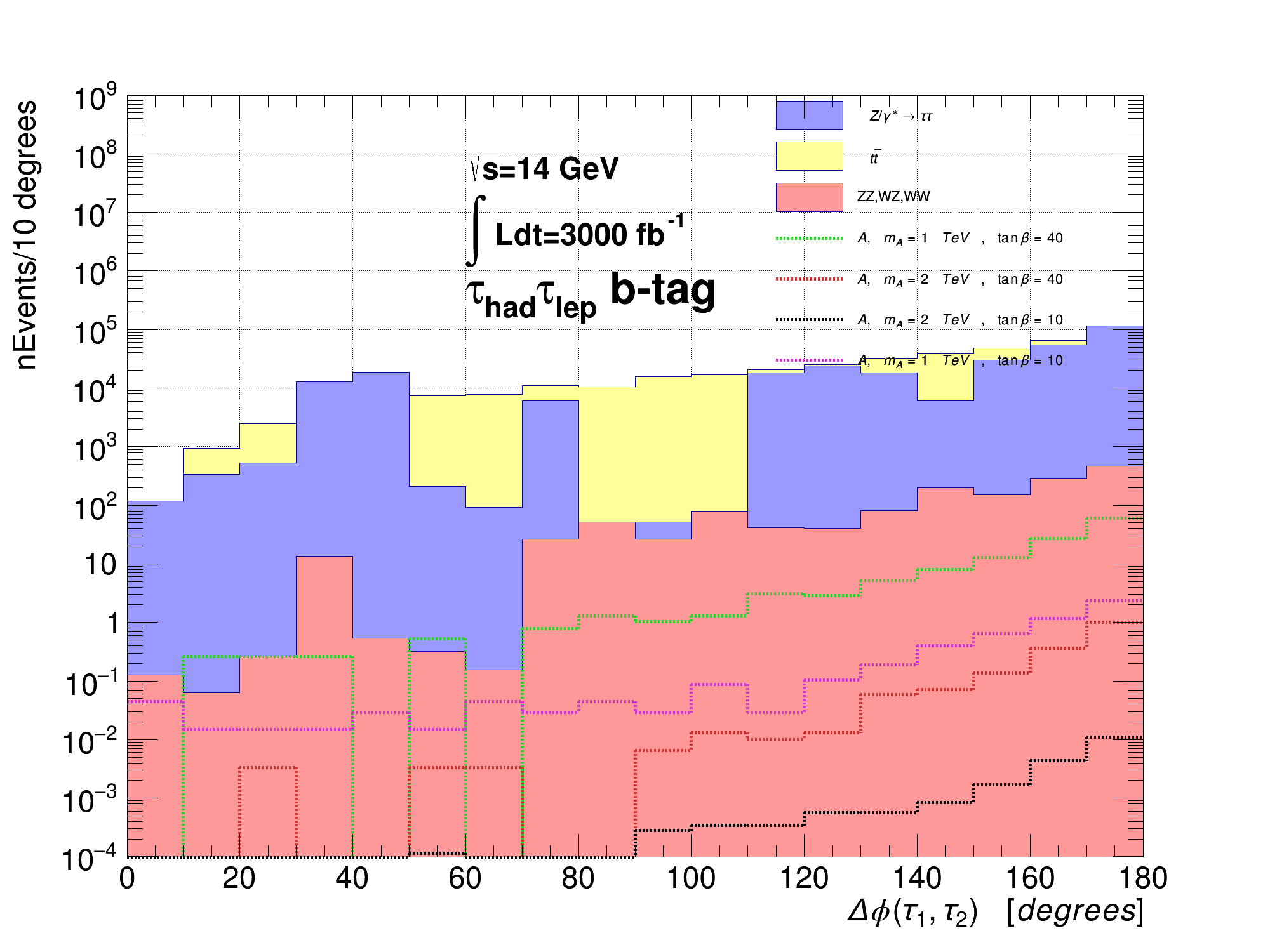}\\
\caption{Distribution in transverse ditau opening angle
  $\Delta\phi (\tau\tau )$ for our $m_h^{125}({\rm nat})$ benchmark
  scenario with $\tan\beta =10$ and $m_A=1$ and 2 TeV.
\label{fig:Dphi}}
\end{center}
\end{figure}

After selecting for candidate ditau events,
we plot in Fig. \ref{fig:Dphi} the transverse opening angle
$\Delta\phi (\tau\bar{\tau} )$ from our signal and BG events for
our $m_h^{125}({\rm nat})$ benchmark point with $m_A=1$ and 2 TeV and
$\tan\beta =10$ and 40. Both the DY background and the signal events
rise to a peak at $180^\circ$ indicating that these events are mostly
back-to-back in the transverse plane as expected. The ditau opening angle from
$t\bar{t}$ and $VV$ are rather less pronounced at $\Delta\phi\sim 180^\circ$.

We next divide our signal into BtB ditau events,
where $\Delta\phi (\tau\bar{\tau} )>155^\circ$ (this Section)
or non-BtB (acollinear) ditaus where $\Delta\phi (\tau\tau )<155^\circ$
(Sec. \ref{sec:mtautau}).

\begin{figure}[htb!]
\begin{center}
\includegraphics[height=0.275\textheight]{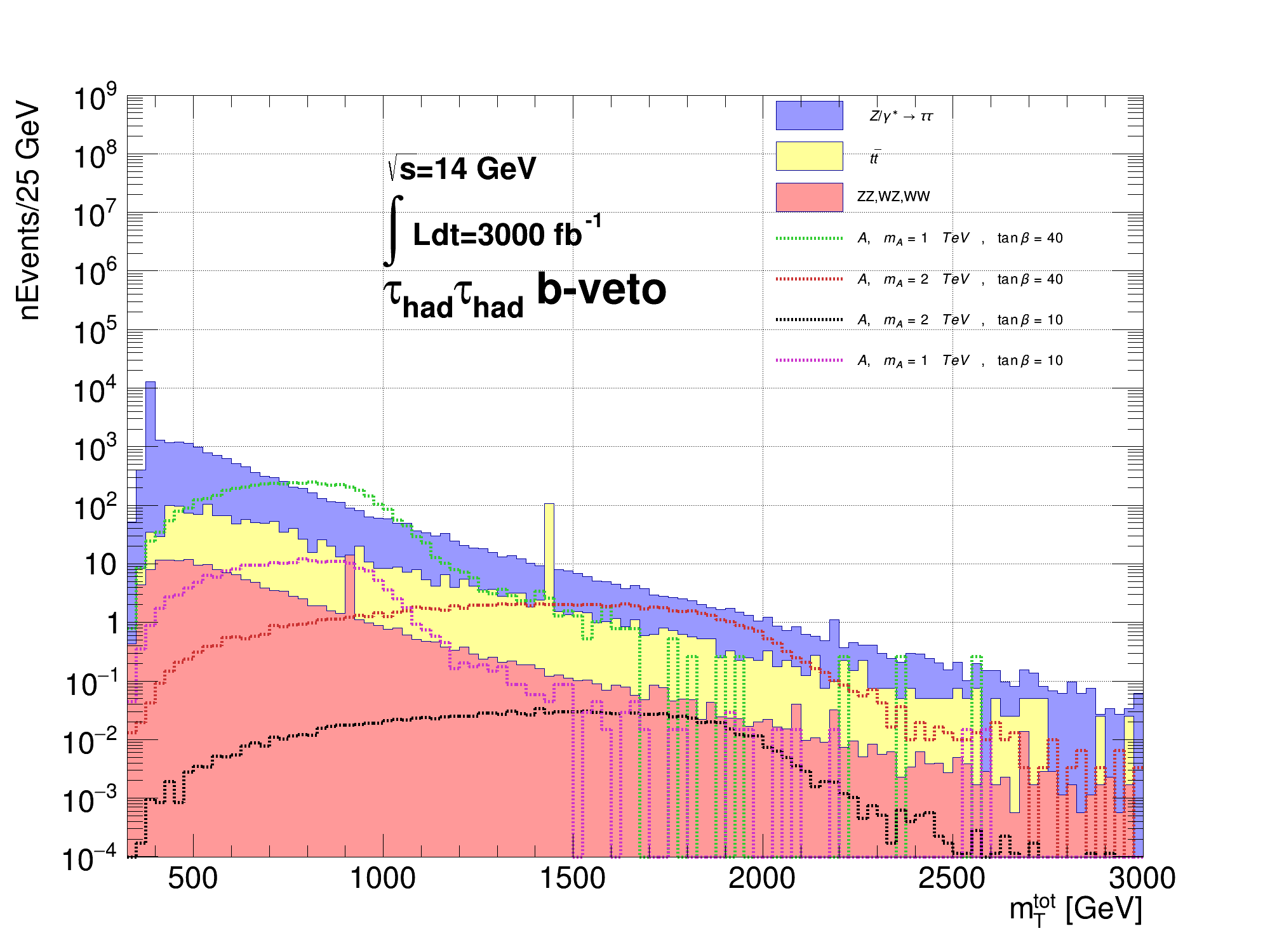}
\includegraphics[height=0.275\textheight]{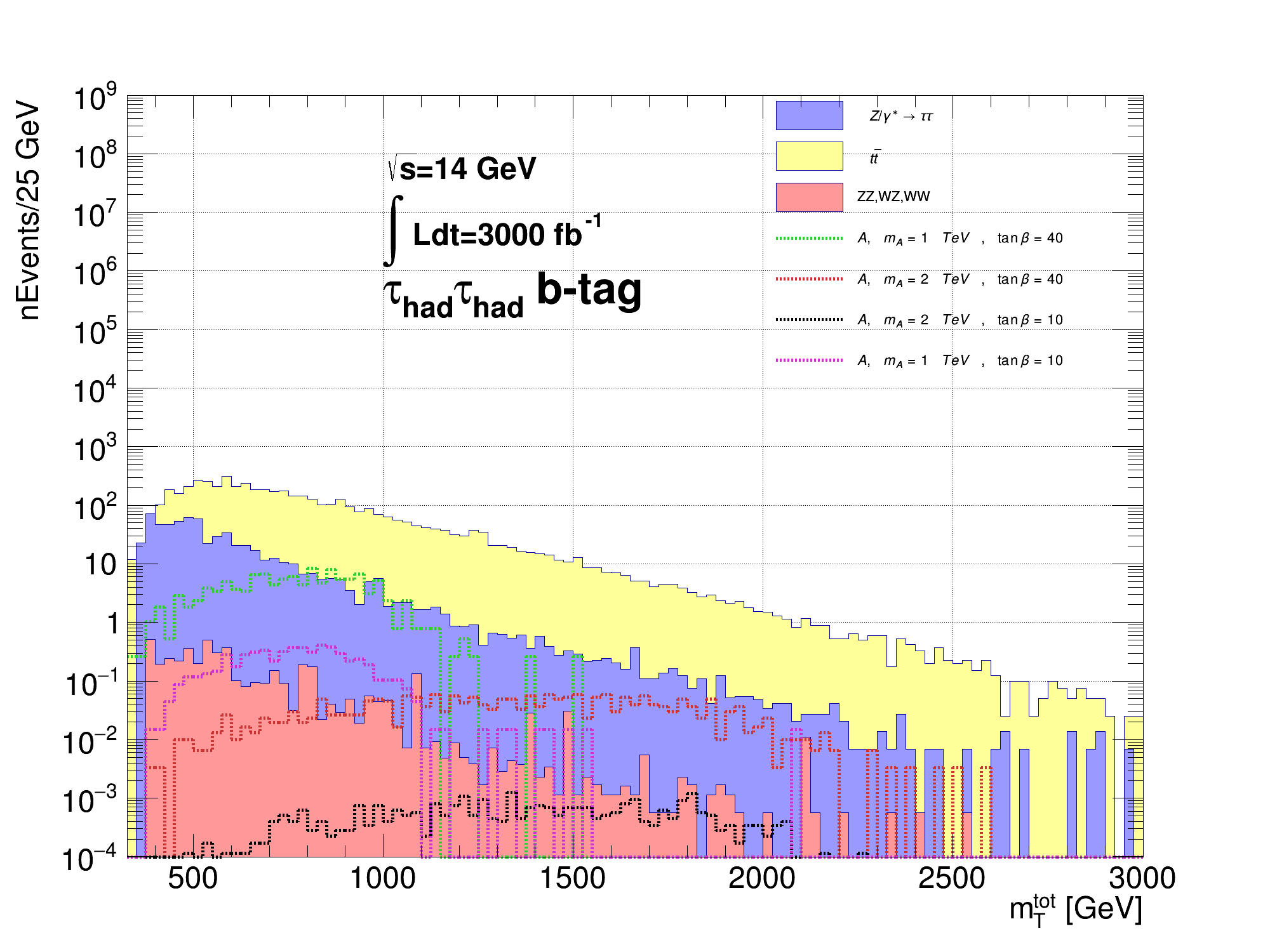}\\
\includegraphics[height=0.275\textheight]{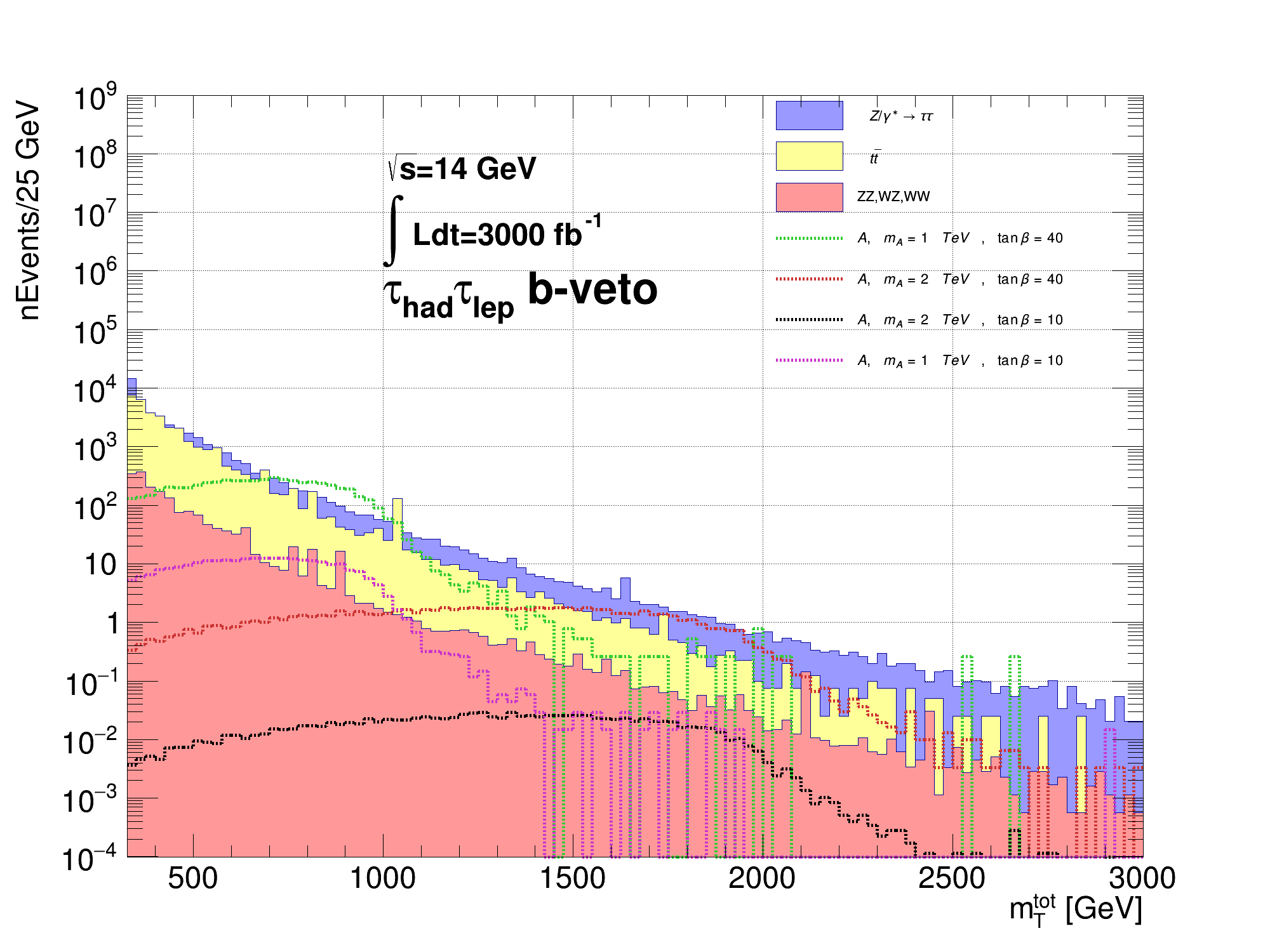}
\includegraphics[height=0.275\textheight]{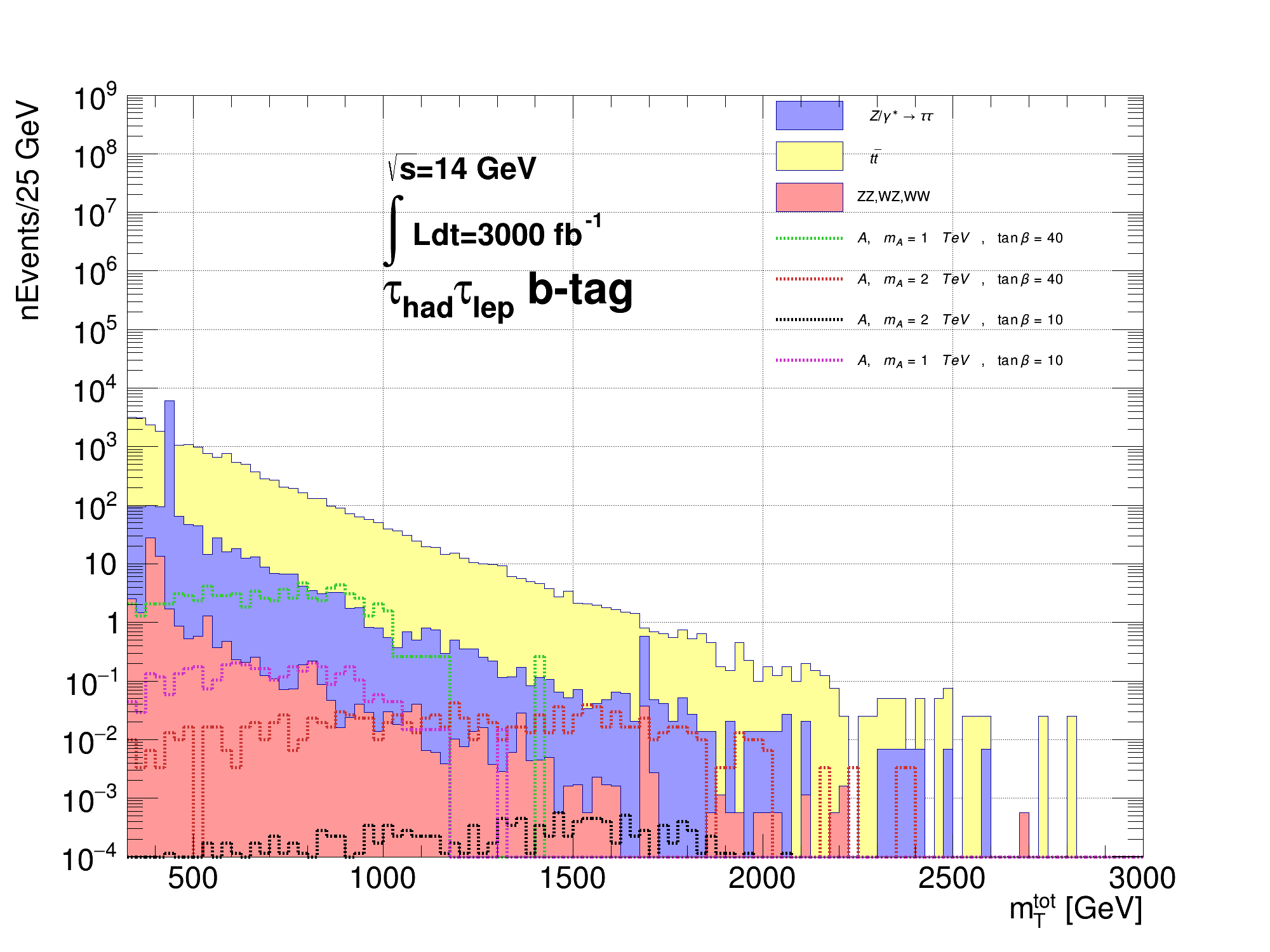}\\
\caption{Distribution in $m_T^{tot}$ for ditau events with
  $\Delta\phi (\tau\tau )>155^\circ$ for our $m_h^{125}({\rm nat})$ benchmark
  scenario with $\tan\beta =10$ and $m_A=1$ and 2 TeV after cuts listed
  in the text.
\label{fig:mT}}
\end{center}
\end{figure}

Then we plot the
total transverse mass variable $m_T^{tot}$ as shown in Fig. \ref{fig:mT}. From the plot, we see that the signal
distributions rise to a peak around $m_T\sim 0.8 m_A$ and then fall off
sharply for $m_T\agt m_A$ due to kinematics
(the cutoff is not completely sharp due to considerable smearing
entering into the signal distributions).
The SM backgrounds are all peaked below $m_T\sim 500$ GeV and have falling
distributions for increasing values of $m_T^{tot}$.

\section{Signal from acollinear $\tau\bar{\tau}$ via $m_{\tau\tau}$}
\label{sec:mtautau}

For acollinear ditau events (non-BtB), we require
the transverse ditau opening angle $\Delta\phi(\tau\tau )<155^\circ$
so that this data set is orthogonal to the back-to-back ditau set.
For the acolliner ditau events, we also require the presence of an
additional jet in the event besides the $\tau_{had}$ jets (usually an initial-state-radiation (ISR)
jet in the case of signal events): $n_{jets}\ge 1$.
For this configuration,
then we are able to use the tau-tau invariant mass reconstruction trick
since once $\vec{E}_T^{miss}$ is known, and we assume the neutrinos from
each tau decay are collinear with the parent tau direction, then the
ditau invariant mass can be solved for.
Since the taus are ultra-relativistic, the daughter visible decay
products and the associated neutrinos are all boosted in the direction of
the parent $\tau$ momentum.
In the approximation that the visibles (vis) and the neutrinos from the
decay of each tau are all exactly collimated in the tau direction,
we can write the momentum carried off by the neutrinos from the decay
$\tau_1\to vis_1\nu$ of the first tau as $\xi_1\vec{p}_T(vis_1)$ and
likewise for the second tau.
Momentum conservation in the transverse plane requires
\be
-\vec{p}_T(j)=(1 + \xi_1)\vec{p}_T(vis_1) + (1 + \xi_2)\vec{p}_T(vis_2).
\ee
Since this is really two independent equations
(recall we require $p_T(j) > 25$ GeV), it is possible
to use the measured values of the jet and visible-tau-decay momenta to
solve these to obtain $\xi_1$ and $\xi_2$, event-by-event.
It is simple to check that in the approximation of collinear tau decay, the
squared mass of the di-tau system is given by
\be
m_{\tau\tau}^2 = (1 +\xi_1)(1 +\xi_2)m_{vis_1 vis_2}^2
\ee
For ditau plus jet events from $H,\ A$-decay to taus, we expect
$\xi_i > 0$ and $m_{\tau\tau}^2$ to peak at $m_{H,\ A}^2$.
Moreover, for these events, the missing energy vector will usually point in
between the two $\tau (vis)$ momentum vectors in the transverse plane.
In contrast, for backgrounds where $E_T^{miss}$ arises from neutrinos from
decays of heavy SM particles ($t$, $W$, $Z$), the visible and $E_T^{miss}$
directions are uncorrelated and the $E_T^{miss}$-vector may point well away,
or even backwards, from one of the leptons so that one (or both)
$\xi_i < 0$.

\begin{figure}[htb!]
\begin{center}
  \includegraphics[height=0.275\textheight]{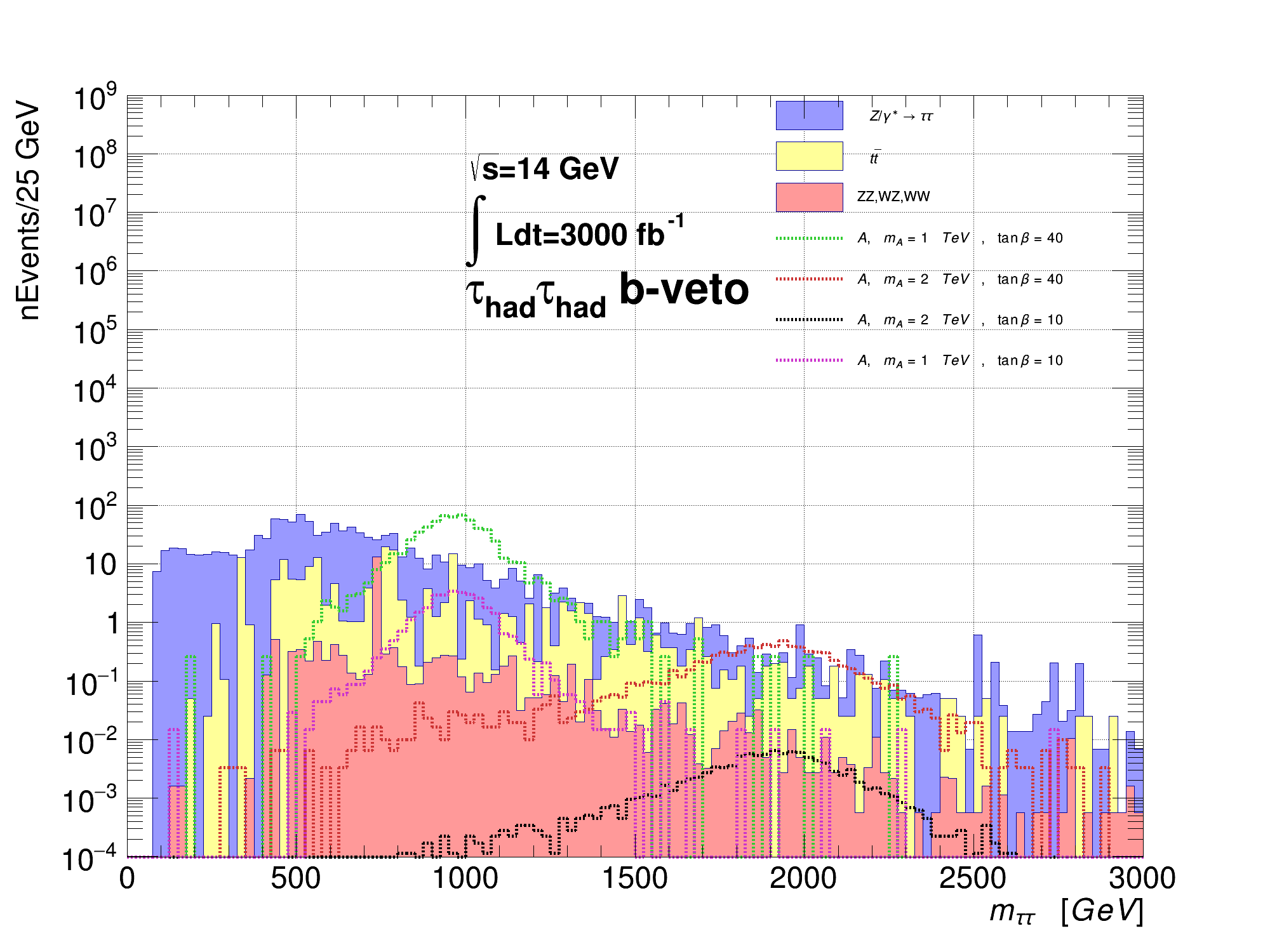}
  \includegraphics[height=0.275\textheight]{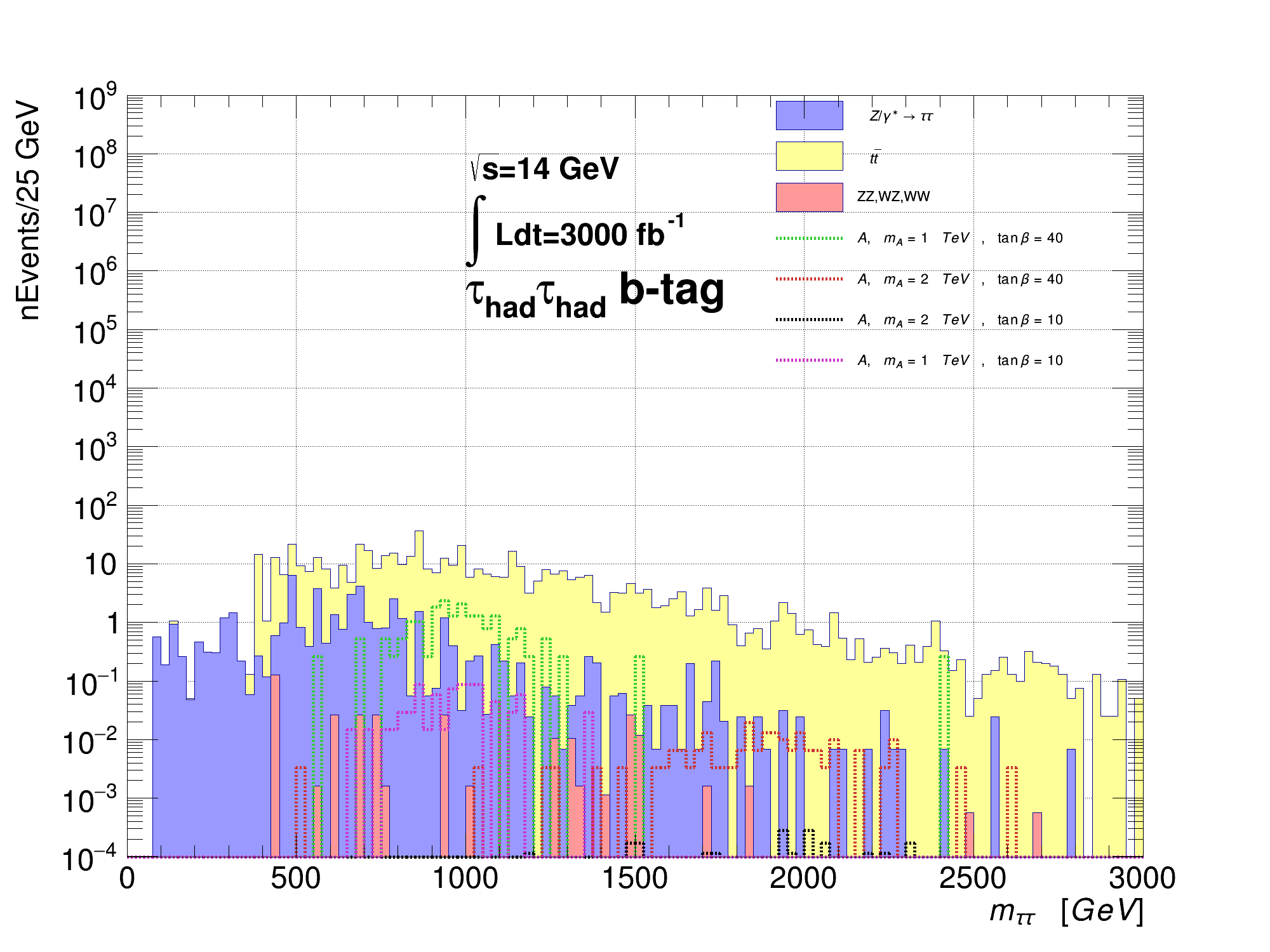}\\
  \includegraphics[height=0.275\textheight]{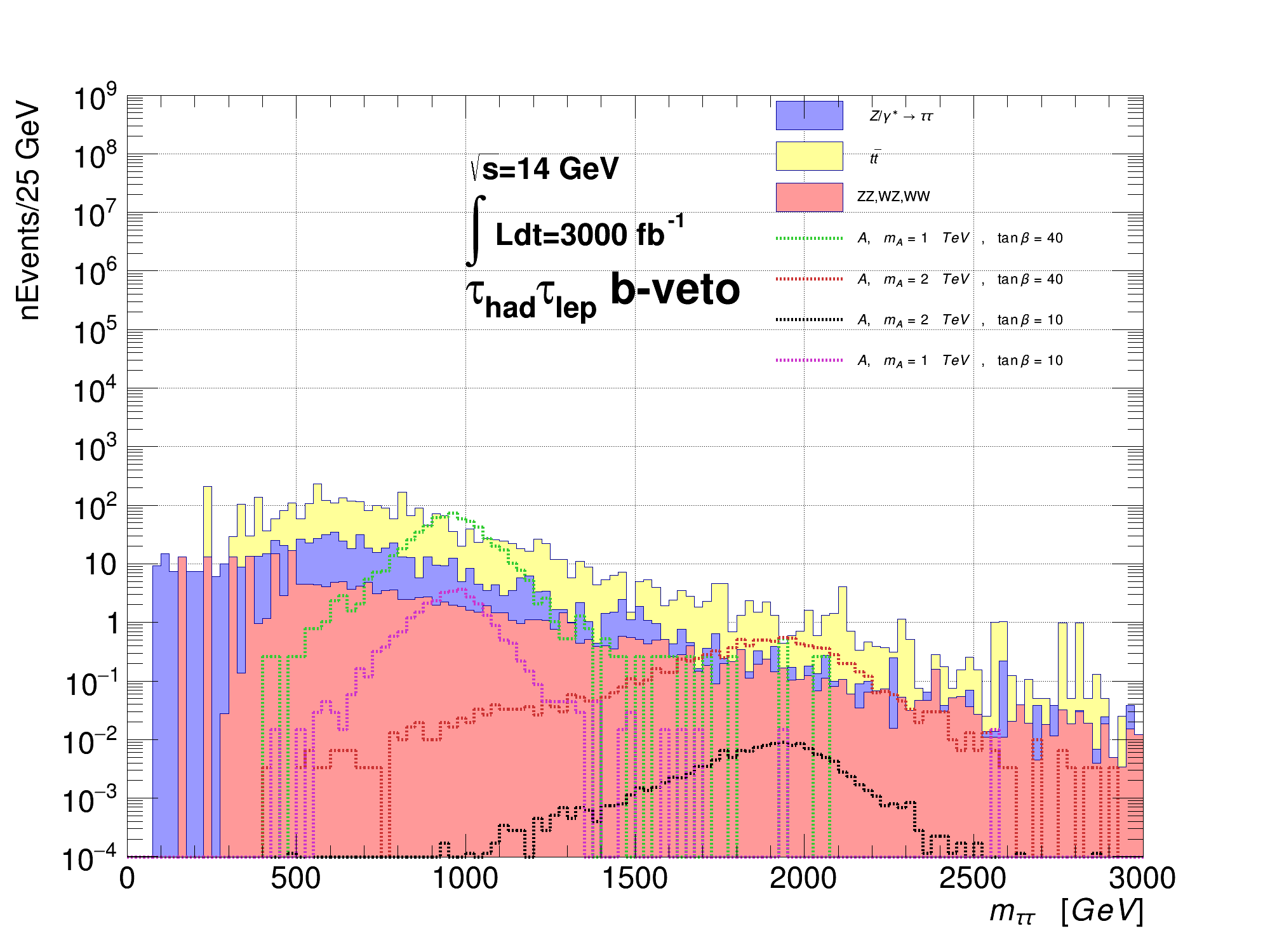}
  \includegraphics[height=0.275\textheight]{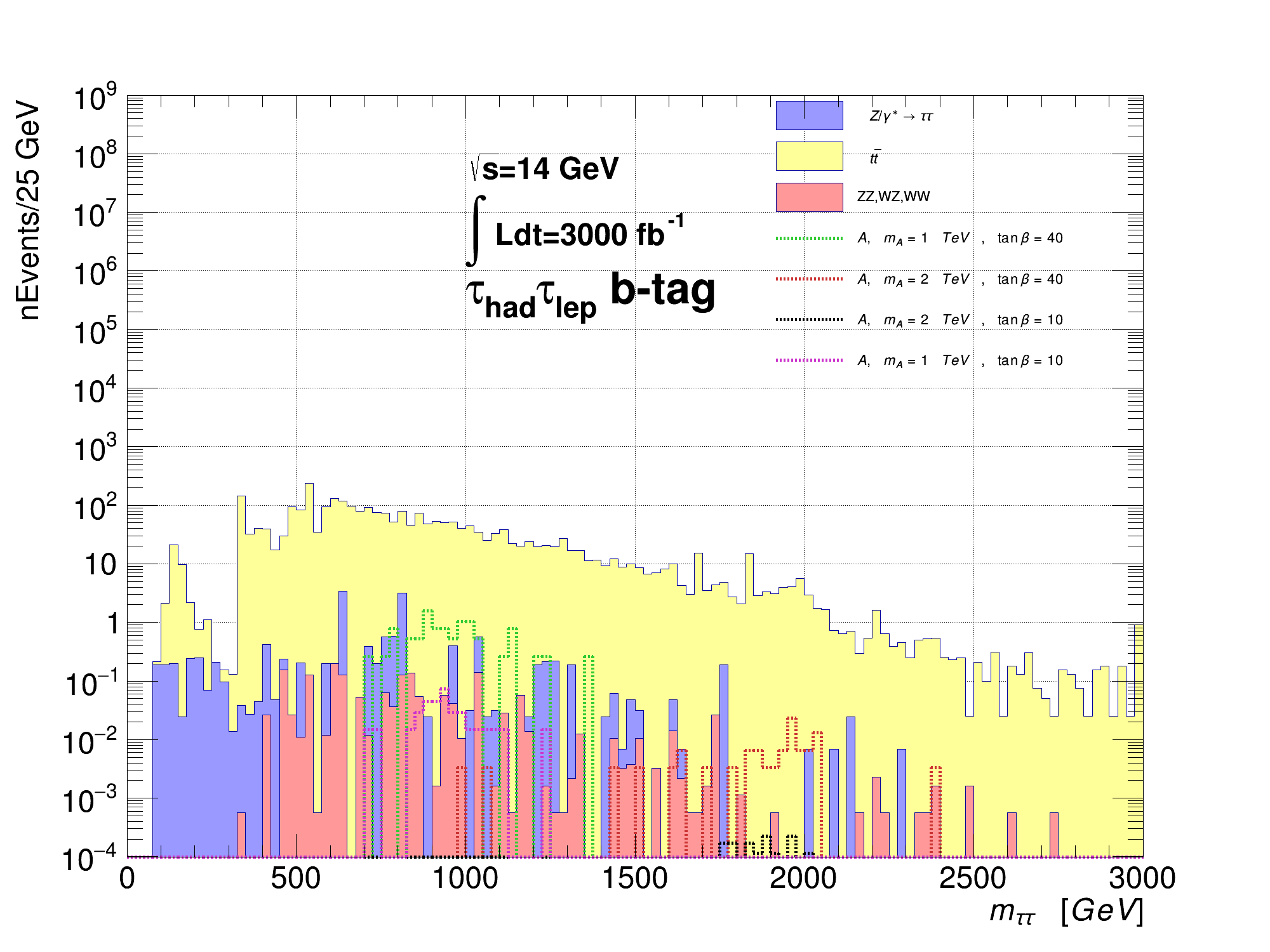}\\
\caption{Distribution in $m_{\tau\tau}$ for ditau events with
  $\Delta\phi (\tau\tau )<155^\circ$ and $n_{jet}\ge 1$
  for our $m_h^{125}({\rm nat})$ benchmark
  scenario with $\tan\beta =10$ and {\it a}) $m_A=1$ and
  {\it b}) $m_A=2$ TeV after cuts listed in the text.
\label{fig:mtautau}}
\end{center}
\end{figure}

Then we can plot the $m_{\tau\tau}$
distribution, as is shown in Fig. \ref{fig:mtautau} for
$\tan\beta =10$ and 40 and for {\it a}) $m_A=1$ TeV and {\it b}) $m_A=2$ TeV.
From the plot, the DY distribution shows a remnant peak at $m_Z=91.2$ GeV
while $t\bar{t}$ and $VV$ are peaked below $500$ GeV. In contrast,
the $A\to\tau\bar{\tau}$ signal distributions are peaked at
$m_{\tau\tau}\sim m_A$ with a width that arises from smearing effects and
non-exact-collinearity of the $\tau$ decay products.

To illustrate some numerics of our results, in Table \ref{tab:sigma}
we list the resultant signal and background cross sections (in fb)
after all cuts for the cases of
$pp\to H,\ A\to\tau\bar{\tau}$ at $\sqrt{s}=14$ TeV for
$\tan\beta =10$ and $m_A=1$ TeV, for both the hMSSM and the $m_h^{125}({\rm nat})$
scenario. From the Table, we see that, as expected, the surviving signal after
cuts from the $m_h^{125}({\rm nat})$ scenario is somewhat diminished
from the hMSSM case due to the diminished branching fractions
$BF(H,\ A\to\tau\bar{\tau})$. Also, the two signal channels
from $H$ and from $A$ production are nearly comparable.
The dominant background comes from $\gamma^*,Z\to\tau\bar{\tau}$ while
$t\bar{t}$ and $VV$ are smaller but still significant. The signal is
quite smaller in the acollinear channel than in the BtB channel. However, this
is compensated for somewhat by smaller backgrounds in the acollinear channel
than in the BtB channel, which makes the acollinear channel to have a much better $S/B$ ratio than the BtB channel.
\begin{table}[h!]
\centering
\begin{tabular}{lcc}
\hline
process & back-to-back (BtB) & acollinear \\
\hline
$H\to\tau\bar{\tau}(hMSSM)$ & 0.197 & 0.024 \\
$A\to\tau\bar{\tau}(hMSSM)$ & 0.222 & 0.027 \\
$H\to\tau\bar{\tau}(SUSY)$ & 0.140 & 0.017 \\
$A\to\tau\bar{\tau}(SUSY)$ & 0.162 & 0.020 \\
\hline
$\gamma^*,Z\to\tau\bar{\tau}$ & 23.33 & 0.586 \\
$t\bar{t}$ & 19.95 & 2.112 \\
$VV$ & 0.663 & 0.069 \\
$total(BG)$ & 43.94 & 2.767 \\
\hline
\end{tabular}
\caption{Cross section (fb) after optimized cuts for the various
  signal and background processes from $pp$ collisions at $\sqrt{s}=14$ TeV
  and $\tan\beta =10$ and $m_A=1$ TeV.
}
\label{tab:sigma}
\end{table}

\section{Reach of LHC3 and HL-LHC for $H,\ A\to\tau\bar{\tau}$}
\label{sec:LHCreach}

After settling on cuts for the BtB and acollinear ditau signals,
it is possible to plot reach plots in terms of exclusion limits or discovery sensitivity for $pp\to H,\ A\to\tau\bar{\tau}$
in the $m_A$ vs. $\tan\beta$ plane. 

For the exclusion plane, the upper limits for exclusion of a signal are set at the 95\% CL and assume the true distribution one observes in experiment corresponds to background only. They are then computed using a modified frequentist $CL_s$ method\cite{Read_2002} with the profile likelihood ratio as the test statistic. 

For the discovery plane, we use $5\sigma$ to denote the discovery and assume the true distribution one observes in experiment corresponds to signal-plus-background. Then we test this against the background only distribution to see if the background only hypothesis could be rejected at a $5\sigma$ level.

In both the exclusion plane and the discovery plane, the asymptotic approximation for getting the median significance is used\cite{Cowan_2011}. The systematic uncertainty is assumed to take $1\sigma$ of the corresponding statistical uncertainty, which is a very conservative rule-of-thumb estimate.

\subsection{Exclusion plane}

\begin{figure}[htb!]
\begin{center}
  \includegraphics[height=0.25\textheight]{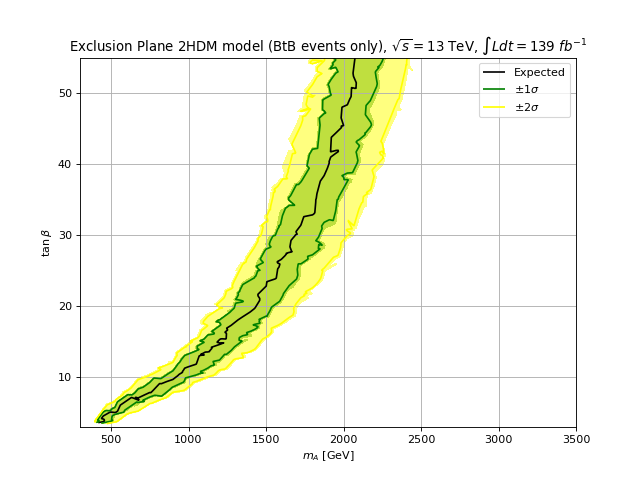}

\caption{The 95\% CL upper limits with $\sqrt{s}=13$ TeV and 139 fb$^{-1}$
  for $H,\ A\to\tau\bar{\tau}$ using BtB signal only in
  the hMSSM.
\label{fig:exl139_btb_hmssm}}
\end{center}
\end{figure}

\begin{figure}[htb!]
\begin{center}
  \includegraphics[height=0.25\textheight]{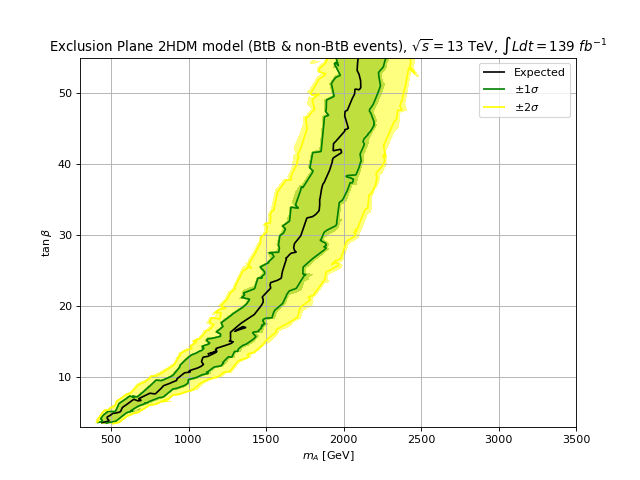}
  \includegraphics[height=0.25\textheight]{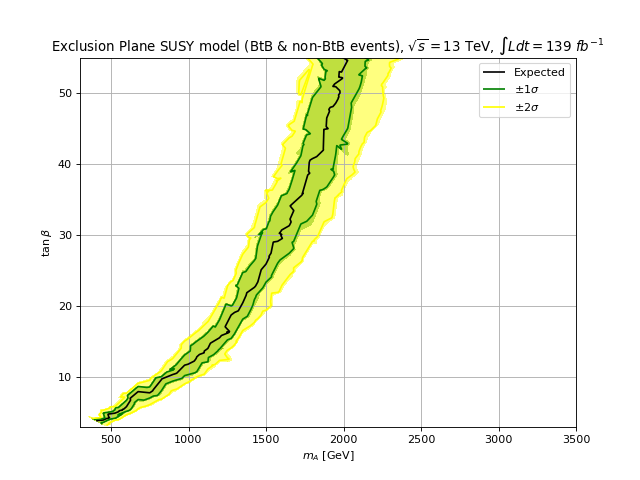}\\
\caption{The 95\% CL upper limits with $\sqrt{s}=13$ TeV and 139 fb$^{-1}$
  for $H,\ A\to\tau\bar{\tau}$ in the
  {\it a}) the hMSSM and {\it b}) the $m_h^{125}({\rm nat})$ scenario.
\label{fig:exl139}}
\end{center}
\end{figure}

As a first step, to compare
with the ATLAS reach of upper limits obtained in their Run 2 search with 139 fb$^{-1}$,
we plot our corresponding exclusion limit in Fig. \ref{fig:exl139_btb_hmssm}. For this plot, we use only the BtB signal in the hMSSM where
$m_h$ is set to 125 GeV, which should compare well with the
$m_h^{125}$ scenario used by ATLAS which contains sparticles at or around 2 TeV,
{\it i.e.} presumably SUSY decay modes are closed for most $m_A$ values
shown in the plot. From Fig. \ref{fig:exl139}, we see our expected
95\% CL exclusion extends to
$m_A\sim 0.9$ TeV for $\tan\beta =10$ which compares favorably with ATLAS. For $\tan\beta =40$, we obtain a 95\% CL exclusion of  $m_A\sim 1.9$ TeV,
which is somewhat better than the ATLAS expected result of $m_A\sim 1.8$ TeV.

In Fig. \ref{fig:exl139}, we plot in frame {\it a}) our expected Run 2 exclusion assuming 139 fb$^{-1}$ using the combined BtB and acollinear signal channels
in the hMSSM. The exclusion limit extends to $m_A\sim 0.95$ TeV 
for $\tan\beta =10$ and to $m_A\sim 1.95$ TeV for $\tan\beta =40$.
For frame {\it b}), for the $m_h^{125}({\rm nat})$ scenario, then the
corresponding 139 fb$^{-1}$ reach extends to $m_A\sim 0.8$ TeV for
$\tan\beta =10$ and to $m_A\sim 1.8$ TeV for $\tan\beta =40$.

\begin{figure}[htb!]
\begin{center}
  \includegraphics[height=0.25\textheight]{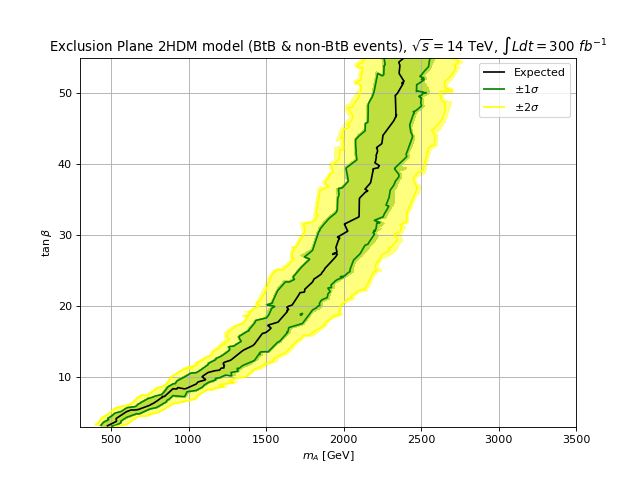}
  \includegraphics[height=0.25\textheight]{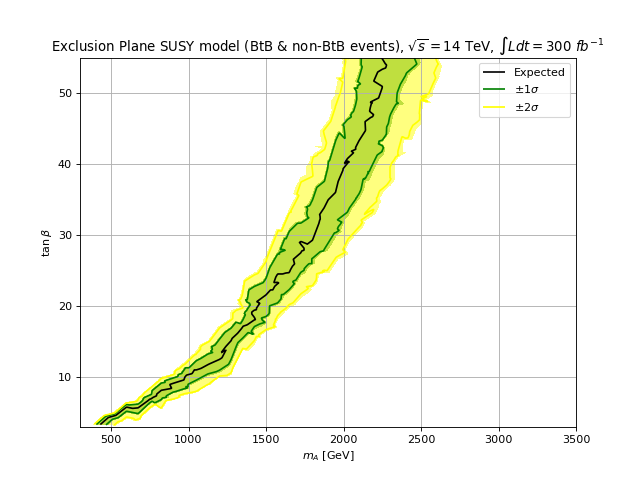}\\
\caption{The 95\% CL upper limits with $\sqrt{s}=14$ TeV and 300 fb$^{-1}$
  for $H,\ A\to\tau\bar{\tau}$ in
  {\it a}) the hMSSM and {\it b}) the $m_h^{125}({\rm nat})$ scenario.
  \label{fig:exl300}}
\end{center}
\end{figure}

In Fig. \ref{fig:exl300}, we present our projected future exclusion plots,
this time for LHC collisions at $\sqrt{s}=14$ TeV with 300 fb$^{-1}$
of integrated luminosity, as would be expected from LHC Run 3.
Here, we use both the BtB and acollinear signals.
For Run 3, we see in frame {\it a}) for the hMSSM with 
$\tan\beta =10$, the 95\% CL exclusion extends out to
$m_A\sim 1.1$ TeV while the $\tan\beta =40$ exclusion extends to
$m_A\sim 2.3$ TeV. For the frame {\it b}) case with the
$m_h^{125}({\rm nat})$ scenario, the 95\% CL reach for $\tan\beta =10$
extends to $m_A\sim 1$ TeV whilst for $\tan\beta =40$ the Run 3
exclusion extends to $m_A\sim 2$ TeV. Thus, comparing the Run 2 139 fb$^{-1}$
exclusion to that expected from LHC Run 3, we find an extra gain in exclusion  of 
$m_A$ of $\sim 0.1-0.2$ TeV. The presence of (natural) SUSY decay modes
tends to reduce the LHC exclusion by $\sim 0.2$ TeV compared to the hMSSM. 

\begin{figure}[htb!]
\begin{center}
  \includegraphics[height=0.25\textheight]{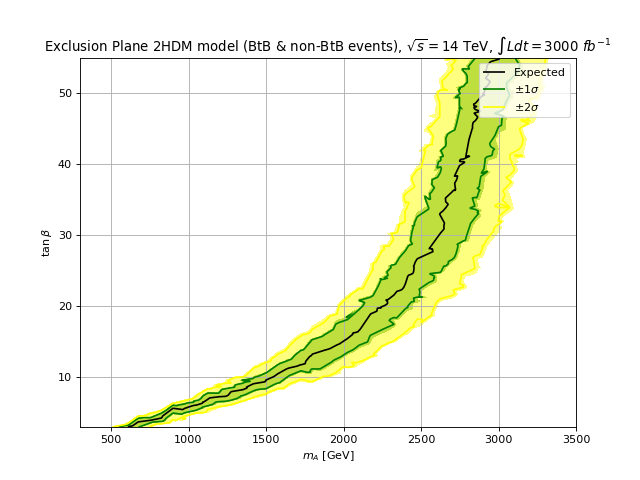}
  \includegraphics[height=0.25\textheight]{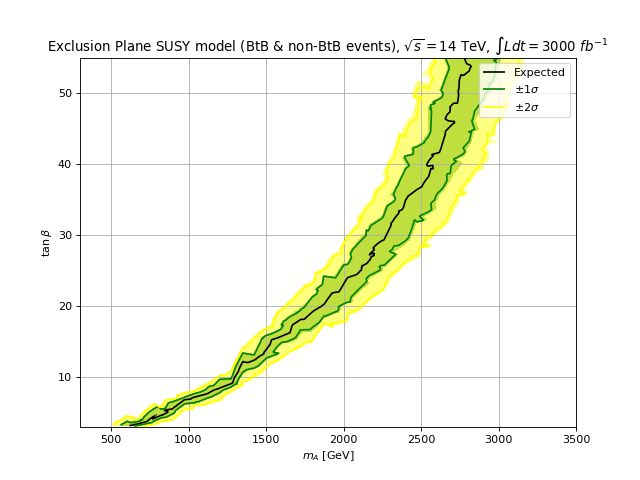}\\
\caption{The 95\% CL upper limits with $\sqrt{s}=14$ TeV and 3000 fb$^{-1}$
  for $H,\ A\to\tau\bar{\tau}$ in
  {\it a}) the hMSSM and {\it b}) the $m_h^{125}({\rm nat})$ scenario.
  \label{fig:exl3000}}
\end{center}
\end{figure}

In Fig. \ref{fig:exl3000}, we plot our projected exclusion limits of HL-LHC
for $H,\ A\to\tau\bar{\tau}$ at $\sqrt{s}=14$ TeV with 3000 fb$^{-1}$.
From frame {\it a}) in the hMSSM case, we find a HL-LHC 95\% CL exclusion out to
$m_A\sim 1.5$ TeV for $\tan\beta =10$ and out to $m_A\sim 2.8$ TeV for
$\tan\beta =40$. If instead we invoke the $m_h^{125}({\rm nat})$
SUSY scenario, then the corresponding HL-LHC exclusion drops to
$m_A\sim 1.3$ TeV for $\tan\beta =10$ and to $m_A\sim 2.6$ TeV
for $\tan\beta =40$, {\it i.e.} a drop in reach of about $0.2$ TeV
in moving from the hMSSM to the $m_h^{125}({\rm nat})$ scenario.

\subsection{Discovery plane}

\begin{figure}[htb!]
\begin{center}
  \includegraphics[height=0.25\textheight]{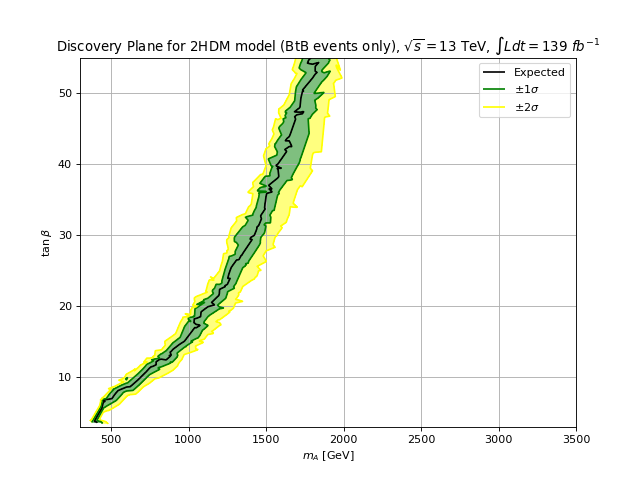}
\caption{The discovery sensitivity at $5\sigma$ level with $\sqrt{s}=13$ TeV and 139 fb$^{-1}$
  for $H,\ A\to\tau\bar{\tau}$ using BtB signal only in
  the hMSSM.
\label{fig:dis139_btb_hmssm}}
\end{center}
\end{figure}

\begin{figure}[htb!]
\begin{center}
  \includegraphics[height=0.25\textheight]{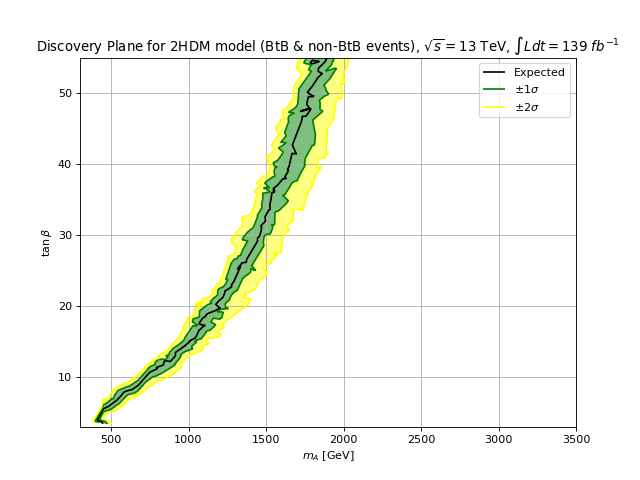}
  \includegraphics[height=0.25\textheight]{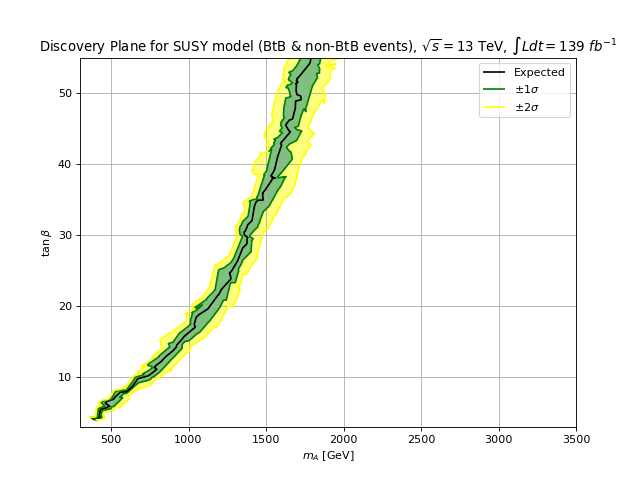}\\
\caption{The discovery sensitivity with $\sqrt{s}=13$ TeV and 139 fb$^{-1}$
  for $H,\ A\to\tau\bar{\tau}$ in the
  {\it a}) the hMSSM and {\it b}) the $m_h^{125}({\rm nat})$ scenario.
\label{fig:dis139}}
\end{center}
\end{figure}

To compare with the ATLAS reach in the discovery plane obtained in their Run 2 search with 139 fb$^{-1}$,
we show our corresponding results in Fig. \ref{fig:dis139_btb_hmssm}. 
For this plot, we use only the BtB signal in the hMSSM where
$m_h$ is set to 125 GeV, which should compare well with the
$m_h^{125}$ scenario used by ATLAS which contains sparticles at or around 2 TeV,
{\it i.e.} presumably SUSY decay modes are closed for most $m_A$ values
shown in the plot. From Fig. \ref{fig:dis139}, we see our expected
$5\sigma$ reach extends to
$m_A\sim 0.75$ TeV for $\tan\beta =10$ which compares favorably wih ATLAS. For $\tan\beta =40$, we obtain a $5\sigma$ reach of  $m_A\sim 1.7$ TeV,
which is somewhat better than the ATLAS expected reach of $m_A\sim 1.6$ TeV.

In Fig. \ref{fig:dis139}, we plot in frame {\it a}) our expected Run 2 discovery reach
assuming 139 fb$^{-1}$ using the combined BtB and acollinear signal channels in the hMSSM. The $5\sigma$ discovery reach for $\tan\beta =10$ extends to $m_A=0.7$ TeV and for $\tan\beta =40$ to $m_A=1.7$ TeV.
For frame {\it b}), for the $m_h^{125}({\rm nat})$ scenario, then the
corresponding 139 fb$^{-1}$ $5\sigma $ discovery reach extends to $m_A\sim 0.7$ TeV for
$\tan\beta =10$ and to $m_A\sim 1.6$ TeV for $\tan\beta =40$.

\begin{figure}[htb!]
\begin{center}
  \includegraphics[height=0.25\textheight]{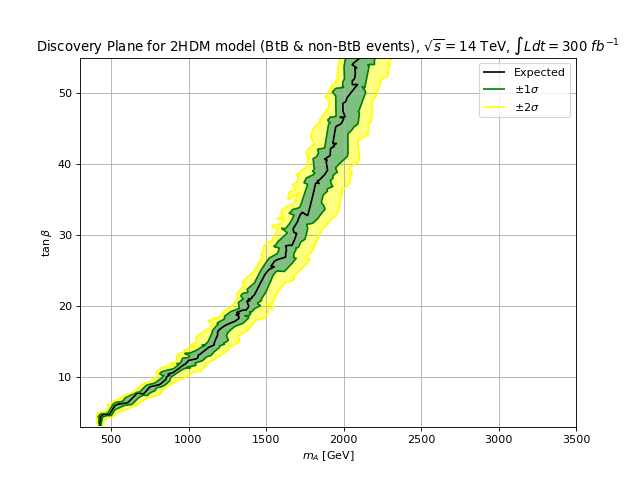}
  \includegraphics[height=0.25\textheight]{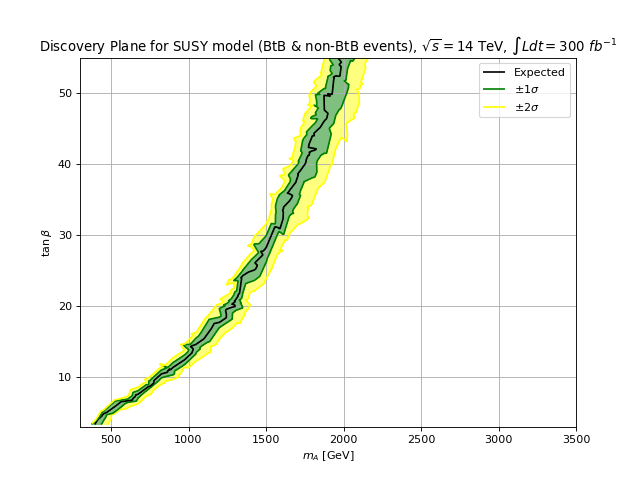}\\
\caption{The discovery sensitivity with $\sqrt{s}=14$ TeV and 300 fb$^{-1}$
  for $H,\ A\to\tau\bar{\tau}$ in
  {\it a}) the hMSSM and {\it b}) the $m_h^{125}({\rm nat})$ scenario.
  \label{fig:dis300}}
\end{center}
\end{figure}

In Fig. \ref{fig:dis300}, we present our future $5\sigma$ discovery sensitivity reach,
this time for LHC collisions at $\sqrt{s}=14$ TeV with 300 fb$^{-1}$
of integrated luminosity, as would be expected from LHC Run 3.
Here, we use both the BtB and acollinear signals.
For Run 3, we see in frame {\it a}) for the hMSSM the
$\tan\beta =10$ discovery reach extends out to
$m_A\sim 0.8$ TeV while the $\tan\beta =40$ reach extends to
$m_A\sim 1.8$ TeV. For the frame {\it b}) case with the
$m_h^{125}({\rm nat})$ scenario, the discovery sensitivity reach for $\tan\beta =10$
extends to $m_A\sim 0.75$ TeV whilst for $\tan\beta =40$ the Run 3
reach extends to $m_A\sim 1.75$ TeV. Thus, comparing the Run 2 139 fb$^{-1}$
reach to that expected from LHC Run 3, we find an extra gain in reach of 
$m_A$ of $\sim 0.1-0.2$ TeV. The presence of (natural) SUSY decay modes
tends to reduce the LHC reach by $\sim 0.1$ TeV compared to the hMSSM. 

\begin{figure}[htb!]
\begin{center}
  \includegraphics[height=0.25\textheight]{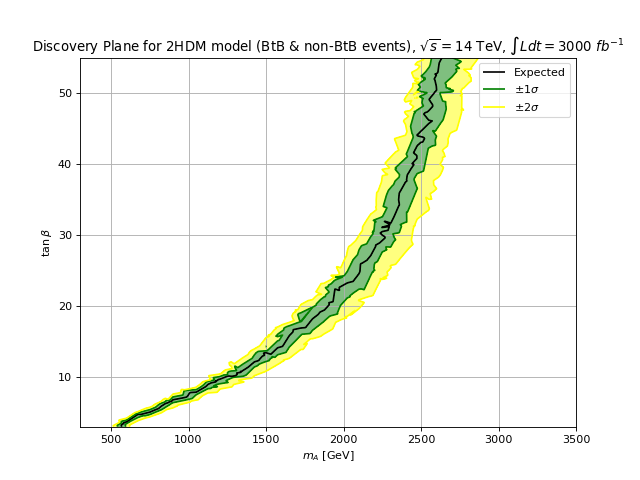}
  \includegraphics[height=0.25\textheight]{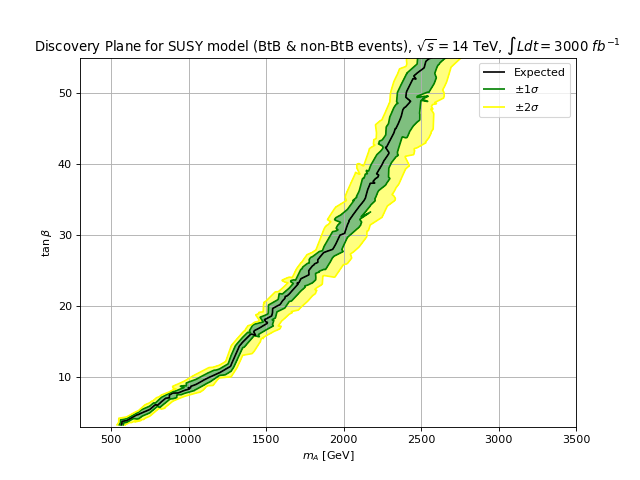}\\
\caption{The discovery sensitivity with $\sqrt{s}=14$ TeV and 3000 fb$^{-1}$
  for $H,\ A\to\tau\bar{\tau}$ in
  {\it a}) the hMSSM and {\it b}) the $m_h^{125}({\rm nat})$ scenario.
  \label{fig:dis3000}}
\end{center}
\end{figure}

In Fig. \ref{fig:dis3000}, we plot our discovery reach of HL-LHC
for $H,\ A\to\tau\bar{\tau}$ at $\sqrt{s}=14$ TeV with 3000 fb$^{-1}$.
From frame {\it a}) in the hMSSM case, we find a HL-LHC discovery sensitivity reach out to
$m_A\sim 1.25$ TeV for $\tan\beta =10$ and out to $m_A\sim 2.45$ TeV for
$\tan\beta =40$. If instead we invoke the $m_h^{125}({\rm nat})$
SUSY scenario, then the corresponding HL-LHC reaches drop to
$m_A\sim 1.15$ TeV for $\tan\beta =10$ and to $m_A\sim 2.25$ TeV
for $\tan\beta =40$, {\it i.e.} a drop in reach of about $0.2$ TeV
in moving from the hMSSM to the $m_h^{125}({\rm nat})$ scenario.

\subsection{Comparing reach results to expectations from the string landscape}

It is instructive to compare the various LHC upgrade reach in $m_A$ to recent
theoretical predictions for SUSY Higgs bosons from the string landscape
picture\cite{Baer:2020kwz}, which also offers a solution to the cosmological constant problem.
In a statistical scan of pocket universes within the greater multiverse as
expected from the string landscape, one expects a power-law draw to large soft
terms\cite{Douglas:2004qg}, including $m_{H_d}^2$ which tends to set the mass scale for $m_{A,H}$.
However, the draw to large soft terms is tempered by the requirement that
contributions to the weak scale should not lie outside the
Agrawal-Barr-Donoghue-Seckel (ABDS) anthropic window\cite{Agrawal:1998xa} lest the weak scale
become too big and complex nuclei and hence atoms as we know them do not form (atomic principle).
In such a setting, the expected statistical predictions in the
$m_A$ vs. $\tan\beta$ plane were plotted in Fig. 9 of Ref. \cite{Baer:2019xww}.
In that Figure, the string landscape with an $n=1$ power-law draw to
large soft terms typically has $m_A$ extending from $1-8$ TeV with
$\tan\beta \sim 10-20$. By comparing our LHC reach plots from either exclusion plane or discovery plane with the string
landscape expectation, we see that even HL-LHC will only probe a small portion
of the theory-expected region of parameter space.

\section{Conclusions}
\label{sec:conclude}

In this paper, we have re-examined the current LHC and LHC-upgrades reach for
SUSY Higgs bosons in a natural SUSY model with $m_h\simeq 125$ GeV.
This led us to propose the $m_h^{125}({\rm nat})$ scenario where a
100 GeV weak scale emerges because all contributions to the weak scale are
comparable to or less than the measured weak scale, in accord with
practical naturalness.
This scenario is a more plausible SUSY benchmark than many
others proposed in the literature in that it requires no implausible finetunings
of parameters in order to gain a value for the weak scale in accord with its measured value. The price of this natural SUSY scenario is that for
$m_A\agt 1-2$ TeV, as is being presently explored at LHC, the
$H,\ A$ decay modes to gaugino+higgsino are frequently open and can even dominate the heavy Higgs branching ratios, thus diluting the
value of the $H,\ A\to\tau\bar{\tau}$ branching fraction as expected in the
hMSSM, or other unnatural SUSY models with a heavy spectrum of SUSY particles.

We also revisited the $H,\ A\to\tau\bar{\tau}$ discovery channels.
Along with the channel used by ATLAS and CMS of BtB ditaus, we
advocated for inclusion of acollinear ditaus where the ditau invariant mass
can be reconstructed under the assumption that the daughter neutrinos
from $\tau$ lepton decay are collinear with the parent $\tau$ direction.
This additional signal channel can substantially increase the signal
compared to using only the BtB ditau channel.

Using the combined BtB and acollinear ditau signals along with the
$m_h^{125}({\rm nat})$ scenario (and the hMSSM for comparison), we
evaluated the present LHC and future LHC upgrades  
exclusion limits and $5\sigma$ discovery reach for
$H,\ A\to\tau\bar{\tau}$ in the $m_A$ vs. $\tan\beta $ plane.
For $\tan\beta =10$, the reach for $m_A$ in the $m_h^{125}({\rm nat})$
senario for Run 2 (Run 3) ((HL-LHC))
extends to $m_A\sim 1$ TeV (1.1 TeV) ((1.4 TeV)).
This will probe some additional chunk of parameter space, although
string landscape  predictions allow $m_A$ values up to $\sim 8$ TeV,
so much higher energy hadron colliders will be needed for a complete
coverage of heavy Higgs boson parameter space.

{\it Acknowledgements:} 

This material is based upon work supported by the U.S. Department of Energy, 
Office of Science, Office of High Energy Physics under Award Number DE-SC-0009956 and DE-SC-001764.

%%%%%%%%%%%%%%%%%%%%%%%%%%%%%%%%%%%%%%%%%%%%%%%%%%%%%%

%\section*{References}
\bibliography{HA}
\bibliographystyle{elsarticle-num}

\end{document}